\documentclass[a4paper,11pt]{article}
\usepackage{pos}
\usepackage{braket}
\usepackage{subcaption}

\title{Multi-hadron interactions from lattice QCD}
\author*[a]{Fernando Romero-L\'opez }

\affiliation[a]{Center for Theoretical Physics, Massachusetts Institute of Technology, Cambridge, MA 02139, USA}

\emailAdd{fernando@mit.edu}

\abstract{
First-principles calculations of multi-hadron dynamics are a crucial goal in lattice QCD. Significant progress has been achieved in developing, implementing, and applying theoretical tools that connect finite-volume quantities to their infinite-volume counterparts. Here, I review some recent theoretical developments and numerical results regarding multi-particle quantities in a finite volume. These results include $N\pi$ scattering, systems of two and three mesons at maximal isospin, three-body resonances in a toy model, and the formulation of effective theories in finite volume for multi-nucleon systems.
}

\FullConference{%
  The 39th International Symposium on Lattice Field Theory (Lattice2022),\\
  8-13 August, 2022 \\
  Bonn, Germany  \\
  Preprint number: MIT/CTP-5516.
}

\begin{document}
\maketitle

\section{Introduction}

Multi-hadron dynamics is a crucial ingredient to understand many important aspects of Quantum Chromodynamics (QCD). Consider first the hadron spectrum, most of the known hadrons are resonances, i.e., unstable particles that appear in multi-body scattering processes.\footnote{See Ref~\cite{Mai:2022eur} for a review.} Examples of this are puzzling excited baryons such as the $\Lambda(1405)$~\cite{Wickramaarachchi:2022mhi}, or the $N(1440)$ (Roper) resonance~\cite{Roper:1964zza}, as well as exotic tetraquark hadrons detected at LHCb~\cite{LHCb:2021vvq,LHCb:2022xob}. 
Similarly, nuclear physics is essentially a multi-nucleon problem, with the binding energies and properties of all atomic nuclei emerging from the interactions of protons and neutrons.

Multi-body interactions also play an important role in weak processes involving hadrons, which are often used to test the Standard Model. Here, I will highlight two examples. First, nucleon resonances can be produced in neutrino-nucleus interactions. Understanding the effects of these resonances is important for reliable predictions of backgrounds in neutrino experiments such as Hyper-Kamiokande or DUNE~\cite{Ruso:2022qes}. Second, several CP violating weak decays of kaons and $D$ mesons have multi-hadron final states, such as $K\to\pi\pi$ (related to the well-known $\epsilon'/\epsilon$ quantity~\cite{NA48:1999szy,KTeV:1999kad}) and $D$ decays to two pions or kaons~\cite{LHCb:2019hro}. The latter are particularly interesting, as no Standard Model prediction is currently available.

Lattice QCD offers the prospect of determining multi-hadron quantities and resonance properties from first principles~\cite{Bulava:2022ovd}. This is an active area of research, as evidenced by the numerous contributions on this topic at the Lattice 2022 conference~\cite{Green:2022rjj,Baeza-Ballesteros:2022bsn,Jackura:2022xml,Severt:2022eic,Wagner:2022bff,Bicudo:2022jep,Padmanath:2022dxy,Riederer:2022apu,Hoffmann:2022jdx,Aoki:2022xxq,Sadl:2022bnq,Bulava:2023mjc,Raposo:2023nex}. More formally, the problem can be formulated as constraining the QCD Scattering Matrix (S-Matrix), $\hat{ \mathcal S}$, from lattice QCD. This object contains all the information about the interactions of the theory in its matrix elements between asymptotic incoming and outgoing states. For instance, resonances emerge as complex poles in $\hat{\mathcal S}$, and the real and imaginary part of the pole position correspond to the resonance mass and width.

However, the determination of multi-hadron quantities differs significantly between experiments and lattice QCD. In experiments, incoming asymptotic states are prepared, interact, and then emerge and become asymptotic states. By measuring the momenta and energy of the products of the different reactions, we can analyze the energy distribution of events or the total cross section. All these quantities are related to the scattering amplitudes and the scattering phases. In contrast, lattice QCD is formulated in Euclidean space-time, so there is no notion of time evolution in the Minkowski sense. Additionally, by working in a finite volume, it is not possible to define asymptotic states in lattice QCD, but only quantum-mechanical stationary states. Lüscher's pioneering work in the 1980s~\cite{Luscher:1986pf,Luscher:1991cf} established a connection between the finite and infinite volume, particularly between the two-particle energy levels and the two-particle scattering amplitude. Enormous progress has been achieved since then, including extensions to three-particle processes and many-body formulations based on effective field theories (EFTs) formulated in finite volume. These theoretical frameworks can be referred to as "finite-volume formalisms".

In this talk, I will present my perspective on the recent progress in multi-hadron quantities from lattice QCD. I will begin with two-hadron processes, then discuss the three-particle formalism and some results, and, finally, ongoing work involving more than three particles.

\section{Two-body interactions in finite volume}

The L\"uscher formalism provides an indirect connection between the energy levels obtained from Euclidean correlation functions and the two-particle scattering amplitude\footnote{See also the HAL QCD method~\cite{Ishii:2006ec,Aoki:2009ji,Aoki:2012tk}.}. Following the original works~\cite{Luscher:1986pf,Luscher:1991cf}, several theoretical extensions have achieved a formalism that enables the study of generic two-to-two systems below the first inelastic threshold with more than two particles~\cite{Rummukainen:1995vs,Kim:2005gf,He:2005ey,Bernard:2010fp,Hansen:2012tf,Briceno:2012yi,Briceno:2014oea,Romero-Lopez:2018zyy,Woss:2020cmp,Grabowska:2021xkp}. The two-particle formalism comes in the form of a determinant equation, the so-called two-particle quantization condition:
\begin{equation}
\det\left[\mathcal{K}_2\left(E^*\right)+F_2^{-1}(E, \boldsymbol{P}, L)\right] \bigg \rvert_{E=E_n}=0,
\end{equation}
where $\mathcal{K}_2(E^*)$ is the two-particle K-matrix that depends on the center-of-mass (CM) energy, and $F_2(E, \boldsymbol{P}, L)$ is a kinematic function, also referred to as the generalized L\"uscher zeta function, which depends on the energy and momentum of the system, as well as the box size $L$. 

This equation offers a beautiful separation of finite-volume effects, contained in $F_2$, and infinite-volume scattering, encoded in $\mathcal K_2$. If a parametrization of the K-matrix is given, for instance in terms of the scattering phase shifts $\delta_\ell$, solutions to the determinant equation correspond to energy levels in finite-volume, $\{ E_n \}$. However, in practice, one typically attempts to solve the inverse problem: given a set of energy levels, what form of the K-matrix best describes the spectrum.

If we consider two scalar particles, and assuming that the lowest partial wave, the $s$ wave, dominates the interactions, the quantization condition provides a one-to-one mapping between each energy level and a value of the $s$-wave phase shift, $\delta_0$. For more complicated systems, this relation does not hold, and one must parametrize the amplitude over some energy range using a functional form based on a few parameters, e.g., an effective range expansion or a phase shift including the Adler zero expected by chiral symmetry~\cite{Blanton:2019vdk}. To obtain the best fit parameters, one would minimize a correlated $\chi^2$ function that quantifies the difference between the lattice QCD energy levels and the predicted ones based on the phase shift.

\subsection{Applications of the two-body formalism}

\begin{figure}[ht]
\begin{subfigure}{.48\textwidth}
  \centering
  % include first image
  \includegraphics[width=\linewidth]{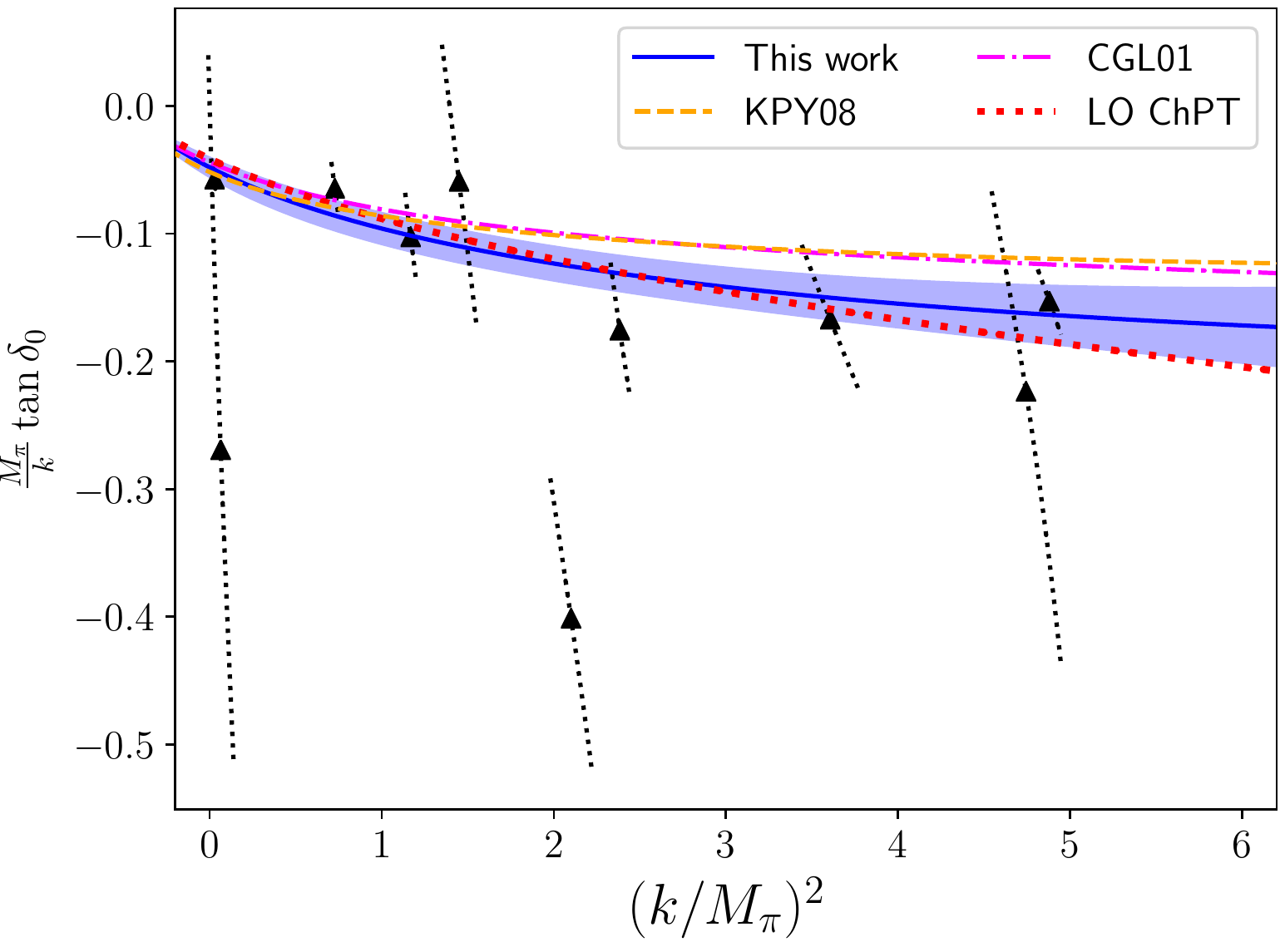}  
  \caption{$I=2$ $\pi\pi$ scattering}
  \label{fig:I2}
\end{subfigure}
\begin{subfigure}{.52\textwidth}
  \centering
  % include second image
  \includegraphics[width=\linewidth]{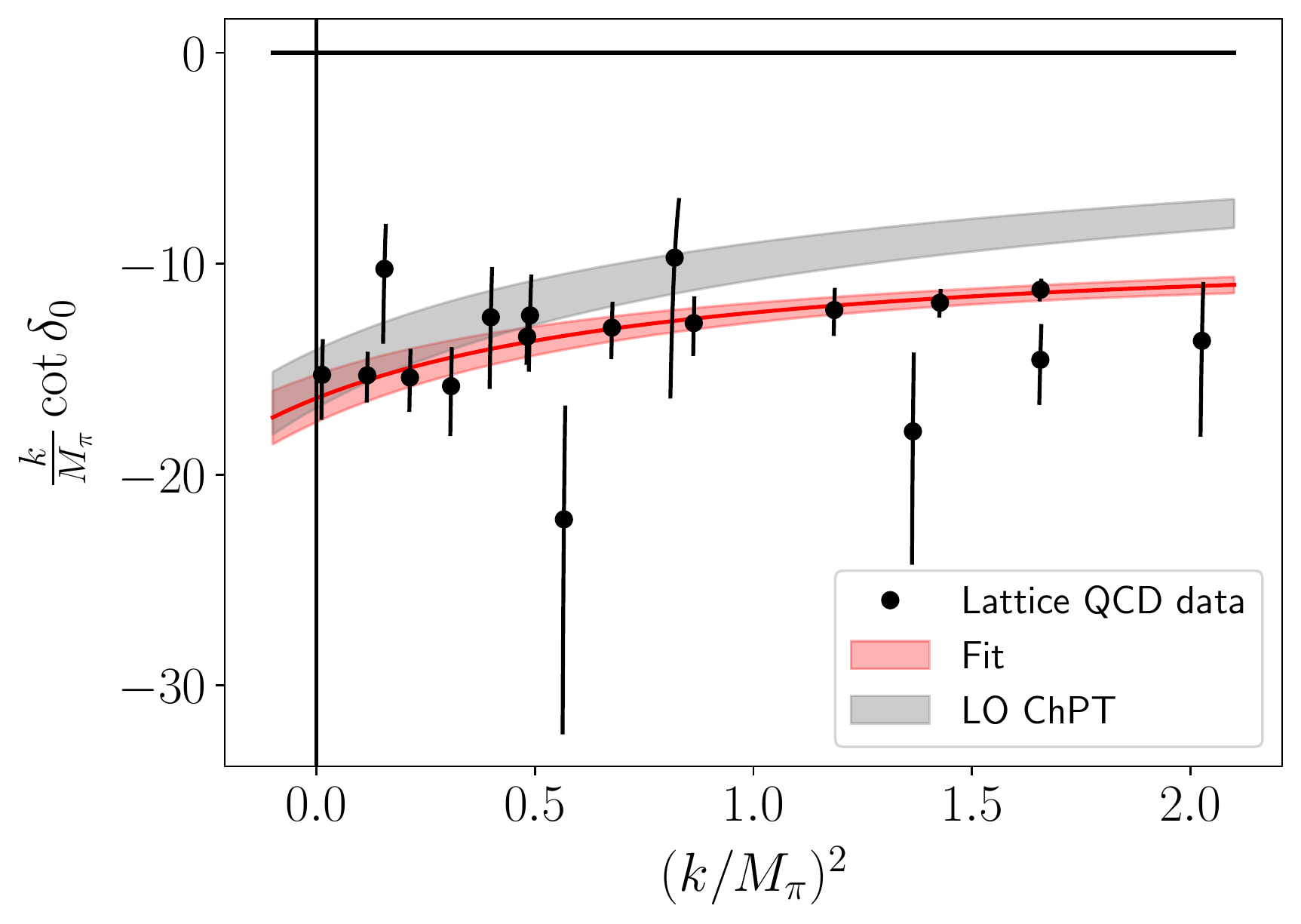}  
  \caption{$I=3/2$ $K\pi$ scattering}
  \label{fig:I32}
\end{subfigure}
\caption{ $s$-wave phase shift, expressed as $(1/k) \tan \delta_0$ or $k \cot \delta_0$, for $\pi\pi$ and $K\pi$  systems at maximal isospin computed from lattice QCD at $M_\pi \sim 140$ MeV. Solid triangles (left) or circles (right) correspond to the phase shift derived from a single energy level. The blue (left) or red (right) band correspond to best fit results for the phase shift, and they are compared against leading order Chiral Perturbation Theory (LO ChPT), and fits to experimental results~\cite{Colangelo:2001df,Kaminski:2006qe}. The gray LO ChPT error band on the right plot is estimated using the variation of the LO prediction with the physical pion or kaon decay constant. The left figure is taken from Ref.~\cite{Fischer:2020jzp}. The right figure involves work in progress~\cite{E250inprog}.}
\label{fig:ppandKp}
\end{figure}

The two-particle quantization condition has been applied to several meson systems, including some at the physical point~\cite{Fischer:2020jzp,Fischer:2020yvw,RBC:2021acc,Bali:2015gji} (see Refs.~\cite{Briceno:2017max,Horz:2022glt} for recent reviews). An example of the latter is given in Fig.~\ref{fig:ppandKp}, where the $s$-wave phase shift for $I=2$ $\pi\pi$ and $I=3/2$ $\pi K$ systems is shown. Since the pion mass is physical, one can expect quantitative agreement with experimental results. This is seen in the left panel of Fig.~\ref{fig:ppandKp}, where the $\pi^+\pi^+$ phase shift agrees within 1-2$\sigma$ with fits to experimental data~\cite{Colangelo:2001df,Kaminski:2006qe}. Note that these lattice results are at a single lattice spacing, and thus, an important future avenue in the topic of meson-meson interactions involves better control over discretization effects. Moreover, Fig.~\ref{fig:ppandKp} corresponds to weakly interacting systems, and it will be interesting to further pursue the study of meson resonances at the physical point, e.g., $\rho$ or $K^*$.

Regarding two-particle systems, meson-baryon or baryon-baryon processes represent the current frontier. These systems are more technically involved, since the signal-to-noise ratio is typically worse than in two-meson systems. Despite this, there has been considerable progress in the last few years in baryon-baryon~\cite{Orginos:2015aya,NPLQCD:2020lxg,Francis:2018qch,Green:2021qol,Horz:2020zvv,Green:2022rjj,Amarasinghe:2021lqa} and meson-baryon~\cite{Fukugita:1994ve,Lang:2012db,Lang:2016hnn,Andersen:2017una,Alexandrou:2017mpi,Verduci:2014btc,Mohler:2012nh,Pittler:2021bqw,Detmold:2015qwf,Torok:2009dg,Bulava:2022vpq} systems. Here, calculations close to the physical point are at an earlier stage, with an article on $N\pi$ scattering at $M_\pi\sim 200$ MeV~\cite{Bulava:2022vpq}, and only preliminary unpublished results for the amplitudes at close-to-physical pion mass~\cite{Pittler:2021bqw}.

\begin{figure}[ht]
  \centering
  % include first image
  \includegraphics[width=\linewidth]{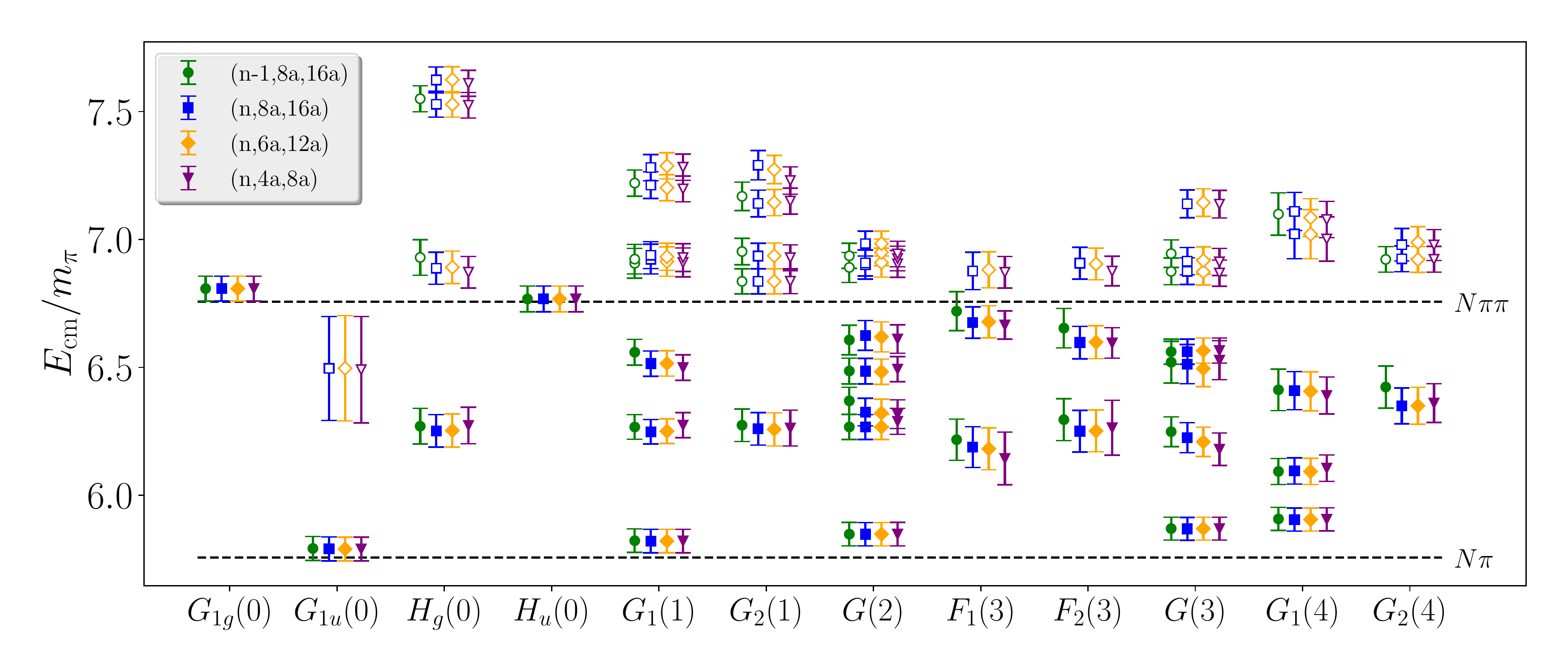}  
  \caption{Energy spectrum in the $I=3/2$ $N\pi$ system at $M_\pi\simeq 200$ MeV from Ref.~\cite{Bulava:2022vpq}. The x axis labels the different irreducible representations (irreps) of the finite-volume symmetry group. Different points correspond to various choices in the parameters of the GEVP. The legend labels $(N_\text{op}, t_0, t_\mathrm{d})$ where $N_\text{op}$ is the number of operators, with $n$ being the maximum number of operators available in each irrep, and $t_0$ and $t_\mathrm{d}$ have been defined in Eq.~\ref{eq:GEVP}.}
  \label{fig:spectrum}
\end{figure}
 
I will now focus on the work of Ref.~\cite{Bulava:2022vpq}. In this and other meson-baryon systems it is important to have reliable extractions of the finite-volume energies. This can be obtained by having an extended set of operators, and solving the generalized eigenvalue problem (GEVP) on a matrix of correlation functions $C(t)$ . One particular choice, referred to as "single pivot", is to first define the eigenvectors $v_n$ of the GEVP as:
\begin{equation}
C\left(t_{\mathrm{d}}\right) v_n\left(t_0, t_{\mathrm{d}}\right)=\lambda_n\left(t_0, t_{\mathrm{d}}\right) C\left(t_0\right) v_n\left(t_0, t_{\mathrm{d}}\right),
\label{eq:GEVP}
\end{equation}
where $t_0$ must be an early time slice, and $t_d$ is an intermediate timeslice in which the diagonalization is performed. Then, the same eigenvectors are used to rotate the corresponding matrix at other time slices $t\neq t_{\mathrm{d}}$. An example of this for the $I=3/2$ $N\pi$ system is shown in Fig.~\ref{fig:spectrum}, where single-baryon ($\Delta$-like) and $N\pi$-like operators are included, and the stochastic Laplacian-Heaviside approach is used~\cite{Peardon:2009gh, Morningstar:2011ka}. The stability of the GEVP is checked by varying the different $t_0$ and $t_{\mathrm{d}}$ choices, as well as by monitoring changes under variations of the operator set. Stable results under these variations provide confidence in the robustness of the spectrum.

\begin{figure}[ht]
  \centering
  % include first image
  \includegraphics[width=0.5\linewidth]{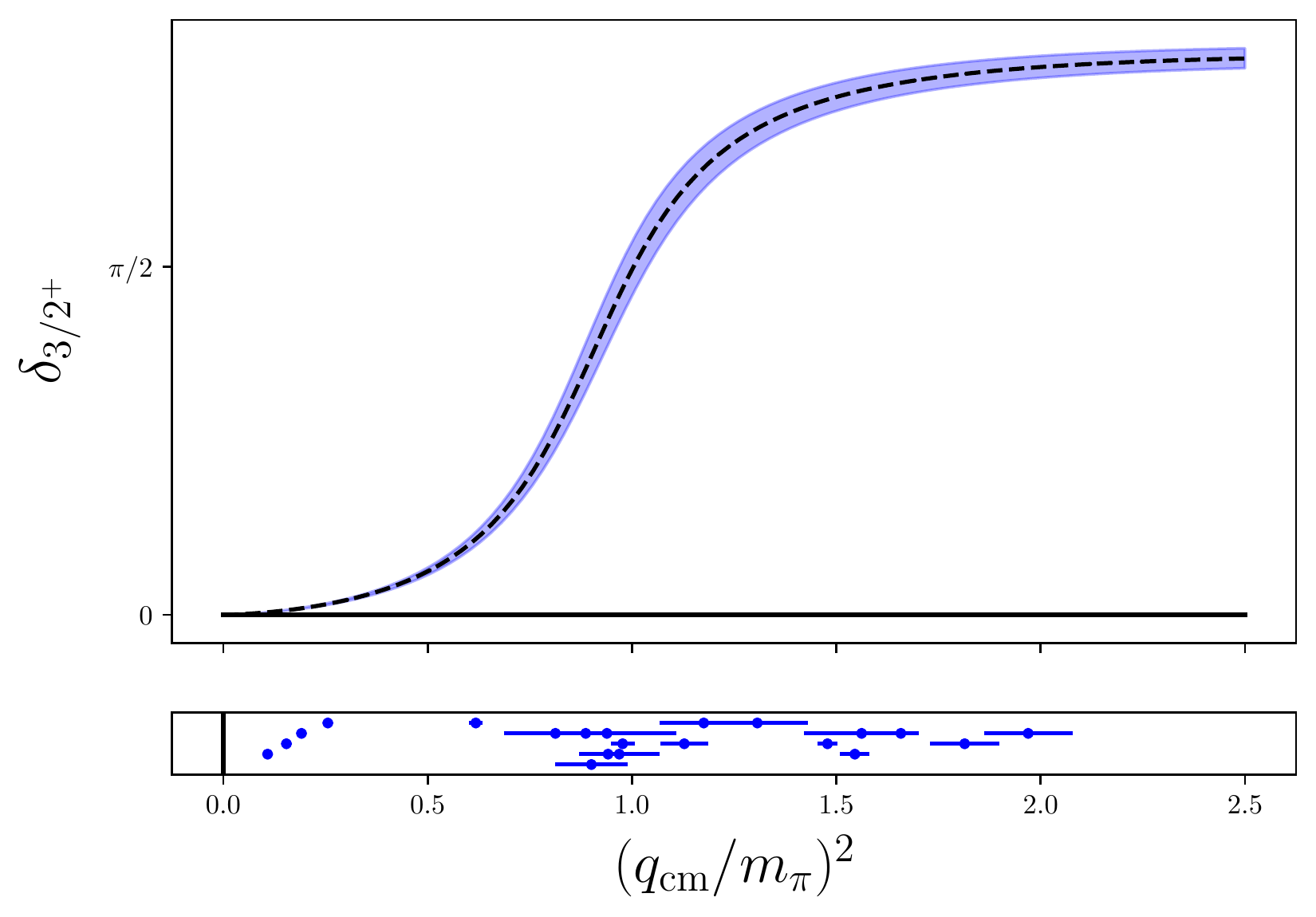}  
  \caption{ Phase shift as a function of the CM momentum for $N\pi$ scattering in the channel of the $\Delta(1232)$ resonance, i.e.~$I=3/2$ and $J^P=3/2^+$. The lower panel shows the position of the energy levels in finite volume used to constrain the phase shift. Figure from Ref.~\cite{Bulava:2022vpq}, with a pion mass of $M_\pi \simeq 200$ MeV.}
  \label{fig:delta1232}
\end{figure}

\begin{figure}[ht]
\begin{subfigure}{.5\textwidth}
  \centering
  % include first image
  \includegraphics[width=\linewidth]{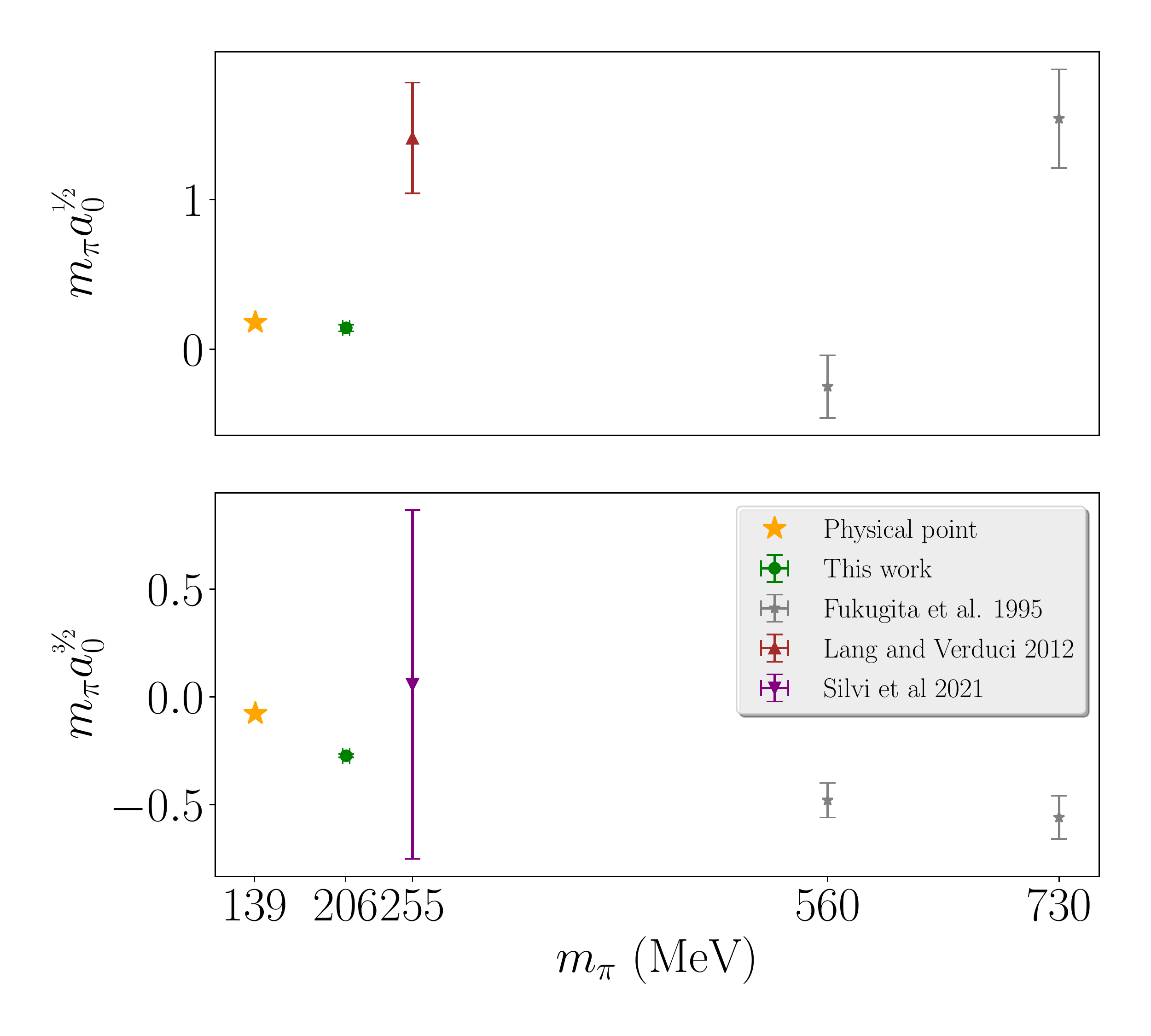}  
  \caption{$N\pi$ $s$-wave scattering lengths.}
  \label{fig:swavepiN}
\end{subfigure}
\begin{subfigure}{.5\textwidth}
  \centering
  % include second image
  \includegraphics[width=\linewidth]{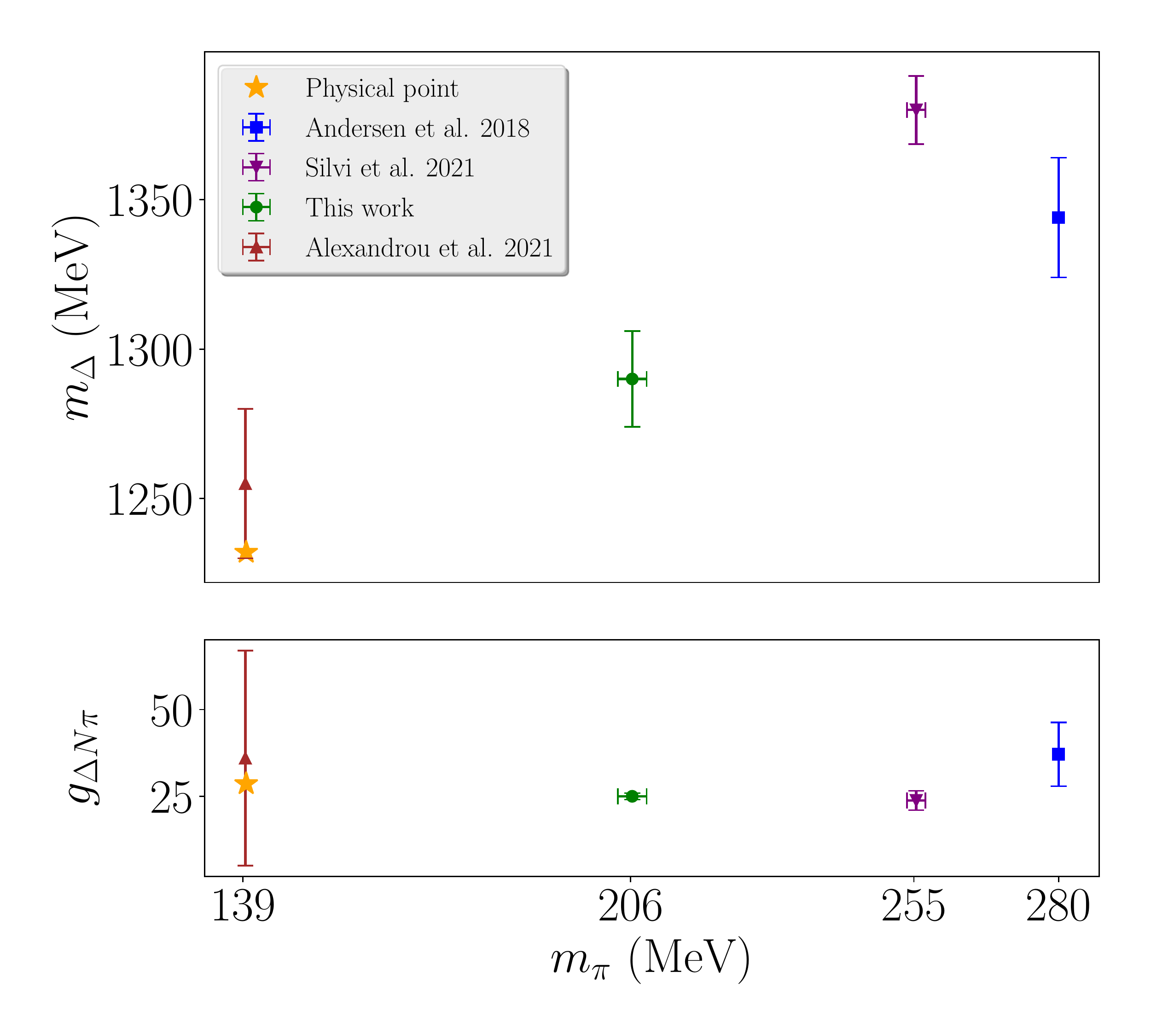}  
  \caption{Breit-Wigner parameters of the $\Delta$ resonance }
  \label{fig:delta}
\end{subfigure}
\caption{ Summary of lattice determinations of the $N\pi$ $s$-wave scattering lengths (left), and the Breit-Wigner parameters of the $\Delta(1232)$ resonance (right). ``This work'' refers to Ref.~\cite{Bulava:2022vpq}, and other results are labeled as ``Anderson et al. 2018'' \cite{Andersen:2017una}, ``Silvi et al. 2021'' \cite{Silvi:2021}, ``Fukugita et al. 1995'' \cite{Fukugita:1994ve}, ``Lang and Verduci 2012'' \cite{Lang:2012db}, and ``Alexandrou et al. 2021''~\cite{Pittler:2021bqw}. Values at the physical point are obtained using Refs.~\cite{pdg:2022,Hemmert:1994ky,Hoferichter:2015tha}. Thanks to S. Skinner for the plot.}
\label{fig:piN}
\end{figure}

The main results of Ref.~\cite{Bulava:2022vpq} are summarized in Figs.~\ref{fig:delta1232} and \ref{fig:piN}. Figure~\ref{fig:delta1232} shows the phase shift of the $\Delta(1232)$. The band shows the best result and uncertainty of the phase shift based on a parametrization via a Breit-Wigner functional form. As can be seen, the phase shift exhibits the behavior of a narrow resonance: a rapid increase crossing $90^{\circ}$ around the position of the resonance. More insight about the comparison to other results is shown in Fig.~\ref{fig:piN}. The right plot compares the Breit-Wigner parameters of the Delta resonance from that work and other compared to the physical point values. There appears to be a somewhat smooth trend towards the physical point result among the different lattice calculations. The left plot shows instead the $s$-wave scattering lengths in the two different isospin channels of an $N\pi$ system. While there is qualitative agreement with the physical point, such as the correct sign in the scattering lengths, the ordering of the $I=1/2$ and $I=3/2$ values is reversed compared to what is observed at the physical point. Another issue is the poor convergence of Chiral Perturbation theory at $M_\pi \sim 200$ MeV. The leading order prediction is significantly different from the values of the scattering length at $M_\pi \simeq 200$ MeV, but at the physical point agrees within 10\% of the result. Further analysis at different values of the pion mass and lattice spacing would be useful in understanding the behavior of chiral EFTs.

\subsection{Matrix elements involving two particles}

To conclude this section, I will discuss another corner of the finite-volume formalism: the computation of matrix elements involving up to two particles. This includes processes such as $K \to \pi\pi$, $\gamma^* \to \pi\pi$ or $\pi \gamma^* \to \pi\pi$ that can be studied via the Lellouch-L\"uscher formalism and its generalizations~\cite{Lellouch:2000pv,Christ:2005gi,Kim:2005gf,Hansen:2012tf,Briceno:2014uqa,Briceno:2015csa,Briceno:2021xlc} (see Refs.~\cite{Feng:2014gba,Andersen:2018mau,Briceno:2015dca,Briceno:2016kkp,Ishizuka:2018qbn,Alexandrou:2018jbt,RBC:2020kdj,Radhakrishnan:2022ubg} for applications). These ideas have been extended to matrix elements of a single current between incoming and outgoing two-particle states~\cite{Baroni:2018iau,Briceno:2019nns,Briceno:2020xxs,Briceno:2020vgp}. Alternatively, we also proposed to study the form factors of resonances by studying the pole position of the resonance as a function of a coupling a static, spatially periodic external field using the L\"uscher formalism in the presence of such a field~\cite{Lozano:2022kfz}. Finally, the formalism for nonlocal matrix elements involving two currents has also been explored in the literature~\cite{delaParra:2020rct,Briceno:2019opb,Briceno:2022omu,Sherman:2022tco}.

\section{Three-particle processes from lattice QCD}

Despite its success, the applicability L\"uscher's two-body formalism is restricted to very few systems in QCD. For instance, many meson resonances, such as the $h_1$ or the $\omega$, have three- or more-pion decay modes. Further examples include the Roper resonance, $N(1440)$, or the doubly-charmed tetraquark, $T_{cc}$ Moreover, the three-neutron force or $NN\Lambda$ interactions can be relevant for an EFT description of the equation of state of neutron stars. Finally, processes with CP violation involve three-meson final states ($K\to 3\pi$) or intermediate states ($K \leftrightarrow 3\pi\leftrightarrow \bar K$).

Extracting three-particle scattering amplitudes from finite-volume calculations is however more complicated than in the two-particle case. This is due, in part, to the fact that three-particle amplitudes can diverge for certain kinematic configurations when the intermediate particle in a one-particle exchange diagram goes on shell. Additionally, three-particle amplitudes contain two-to-two subprocesses, and so, they depend on two-particle interactions. Nevertheless, significant progress has been made in understanding the three-particle problem in finite volume, as can be seen in recent reviews~\cite{Hansen:2019nir,Rusetsky:2019gyk,Mai:2021lwb,Romero-Lopez:2021zdo}. These advances include theoretical developments~\cite{Polejaeva:2012ut,Hansen:2014eka,Hansen:2015zga,Hansen:2015zta,Hansen:2016ync,Briceno:2017tce,Briceno:2018mlh,Briceno:2018aml,Blanton:2019igq,Briceno:2019muc,Jackura:2019bmu,Romero-Lopez:2019qrt,Hansen:2020zhy,Blanton:2020gmf,Blanton:2020jnm,Blanton:2020gha,Hansen:2021ofl,Blanton:2021mih,Blanton:2021eyf,Hammer:2017uqm,Hammer:2017kms,Doring:2018xxx,Romero-Lopez:2018rcb,Pang:2019dfe,Romero-Lopez:2020rdq,Muller:2020wjo,Muller:2020vtt,Mai:2017bge,Guo:2017ism,Klos:2018sen,Guo:2018ibd,Guo:2019hih,Pang:2020pkl,Jackura:2020bsk,Muller:2021uur,Jackura:2022gib}, as well as the first applications to lattice QCD data~\cite{Mai:2018djl,Blanton:2019vdk,Mai:2019fba,Culver:2019vvu,Guo:2020kph,Fischer:2020jzp,Alexandru:2020xqf,Hansen:2020otl,Brett:2021wyd,Mai:2021nul,Blanton:2021llb,Blanton:2021eyf} boosted by technical developments in the extractions of the lattice spectrum~\cite{HadronSpectrum:2009krc,Morningstar:2011ka,Horz:2019rrn}. In this section, I will summarize some of these results.

\subsection{Overview of the formalism}
\label{sec:qc3}

The three-particle formalism was derived in three different ways: using (i) a generic relativistic effective field theory (RFT)~\cite{Hansen:2014eka,Hansen:2015zga}, (ii) a nonrelativistic effective field theory (NREFT)~\cite{Hammer:2017uqm,Hammer:2017kms}, and (iii) the (relativistic) finite volume unitarity (FVU) approach~\cite{Mai:2017bge,Mai:2018djl}.
Initially, all three approaches dealt with identical (pseudo)scalars with $G$-parity-like symmetry, e.g.~a $3\pi^+$ system in isospin-symmetric QCD. Subsequent developments have extended the formalism to distinguishable particles~\cite{Hansen:2020zhy,Blanton:2021mih,Blanton:2020gmf} and $2 \to 3$ transitions~\cite{Briceno:2017tce}. Work is currently underway to include three particles with spin~\cite{3neutrons}. The three different versions of the finite-volume formalism are equivalent. The relation between the RFT and FVU approaches has been established~\cite{Blanton:2020jnm,Jackura:2022gib}, and in the limit of only $s$-wave interactions, the expressions for the FVU and the NREFT in its relativistic formulation~\cite{Muller:2021uur} are identical.

For the sake of concreteness, I will focus on the RFT approach, for which the three-particle quantization condition for identical (pseudo)scalars is~\cite{Hansen:2014eka}
\begin{equation}
\det \left[\mathcal K_\text{df,3}(E^*) + F_3^{-1}(E,{\boldsymbol P}, L)\right] \bigg \rvert_{E=E_n} = 0. \label{eq:QC3}
\end{equation}
Here, all objects are matrices in a space that describes three on-shell particles: two of the particles  (interacting pair) with angular momentum indices, $\ell m$, and a third particle (spectator) labeled by finite-volume three-momentum, $\boldsymbol k$. The matrix $F_3$ depends on kinematics as well as on the two-particle $K$-matrix, $\mathcal K_2$:
\begin{equation}
F_3 = \frac{F_2}{3} - F_2 \frac{1}{ 1/\mathcal K_2+ F_2 + G} F_2, \label{eq:F3}
\end{equation}
where $F_2$ and $\mathcal K_2$ are substantially the same as in the two-particle quantization condition, and $G$ accounts for the finite-volume effects arising from one-particle exchange diagrams.

$\mathcal{K}_\text{df,3}$ is the quantity that parametrizes three-body effects in Eq.~\ref{eq:QC3}. It is, however, unphysical due to the presence of scheme (or cutoff) dependence. The relation to the physical three-particle scattering amplitude, $\mathcal M_3$, requires solving integral equations that map the K-matrices onto $\mathcal M_3$~\cite{Hansen:2015zga}. Schematically, the procedure is:
\begin{equation}
\mathcal{K}_2, \mathcal{K}_\text{df,3} \xrightarrow[\text{ Integral equations }]{} \mathcal M_3.
\end{equation}
Solutions to integral equations have been explored in a series of works~\cite{Briceno:2018mlh,Hansen:2020otl,Jackura:2020bsk,Garofalo:2022pux}.

Generalizations of the RFT approach to nonidentical particles~\cite{Hansen:2020zhy,Blanton:2021mih,Blanton:2020gmf} are fundamentally very similar to the formalism just described. More specifically, all the objects are promoted to matrices in a space with an additional index labeling the flavor of the spectator particle.

To conclude, it is also worth mentioning that the formalism to study matrix elements involving three particles has been explored. In particular, for one-to-three decays such as $K\to3\pi$~\cite{Muller:2020wjo,Hansen:2021ofl,Muller:2022oyw}, as well as for nonlocal matrix elements involving two currents, relevant for $K-\bar K$ mixing~\cite{Jackura:2022xml}.

\subsection{Applications to weakly-interacting systems}

The computation of $\mathcal K_\text{df,3}$ from lattice QCD is one of the main goals of the three-particle formalism. In the following, I will discuss some efforts to determine $\mathcal K_\text{df,3}$ for weakly interacting systems consisting of $\pi^+$ and/or $K^+$. These are the natural starting point, as they represent the simplest systems in QCD.

Let us start with the case of identical particles. In order to analyze the three-particle spectrum, it will be important to parametrize $\mathcal K_\text{df,3}$ around the three-particle threshold. This can be performed by utilizing that $\mathcal K_\text{df,3}$ is a smooth function in some region about threshold of Lorentz-invariant combinations of Mandelstam variables that have the correct symmetries: parity, time-reversal and invariance under particle exchange. As explained in Ref.~\cite{Blanton:2019igq}, to quadratic order, the threshold expansion of $\mathcal K_\text{df,3}$ has only five free parameters:
\begin{equation}
\mathcal{K}_{\mathrm{df}, 3}  =\mathcal{K}_{\mathrm{df}, 3}^{\text {iso }, 0}+\mathcal{K}_{\mathrm{df}, 3}^{\text {iso } 1} \Delta+\mathcal{K}_{\mathrm{df}, 3}^{\text {iso }, 2} \Delta^2  +\mathcal{K}_A \Delta_A+\mathcal{K}_B \Delta_B+O\left(\Delta^3\right) \label{eq:Kdfdwave}
\end{equation}
where $\mathcal{K}_\text{df,3}^{\mathrm{iso}, 0}, \mathcal{K}_\text{df,3}^{\mathrm{iso}, 1}, \mathcal{K}_\text{df,3}^{\mathrm{iso}, 2}, \mathcal{K}_{A}, \mathcal{K}_{B}$ are real constants, $\Delta = (s-9M_\pi^2)/(9M_\pi^2)$ with $s$ being the square of the CM energy of the three particle system, and $\Delta_{A/B}$ are also kinematic functions of Mandelstam variables---see e.g. Eqs. 2.15 and 2.16 in Ref.~\cite{Blanton:2019igq}.

\begin{figure}[ht]
\begin{subfigure}{.5\textwidth}
  \centering
  % include first image
  \includegraphics[width=\linewidth]{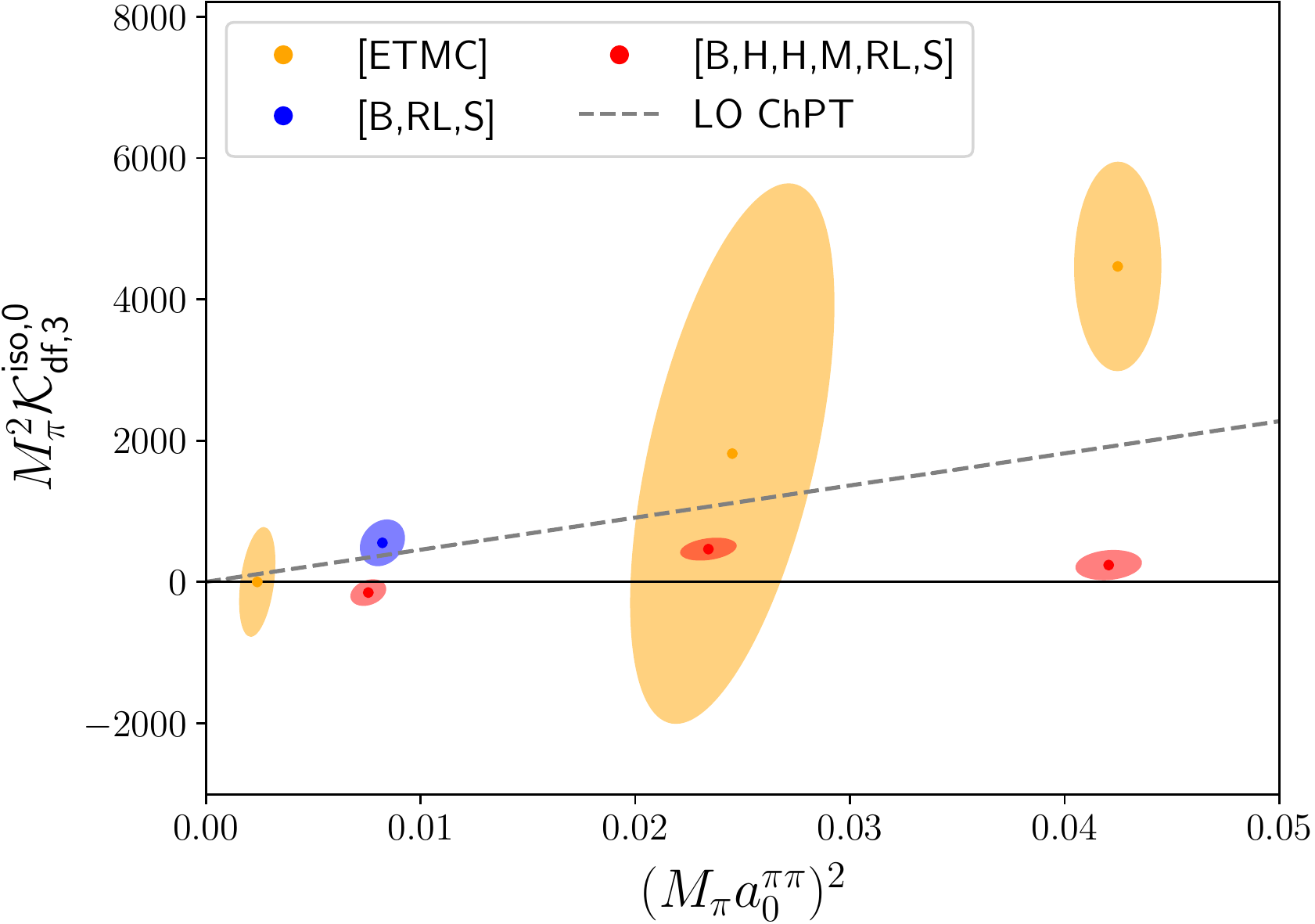}  
  \caption{$\mathcal{K}_{\mathrm{df}, 3}^{\text {iso }, 0}$}
  \label{fig:K0}
\end{subfigure}
\begin{subfigure}{.5\textwidth}
  \centering
  % include second image
  \includegraphics[width=\linewidth]{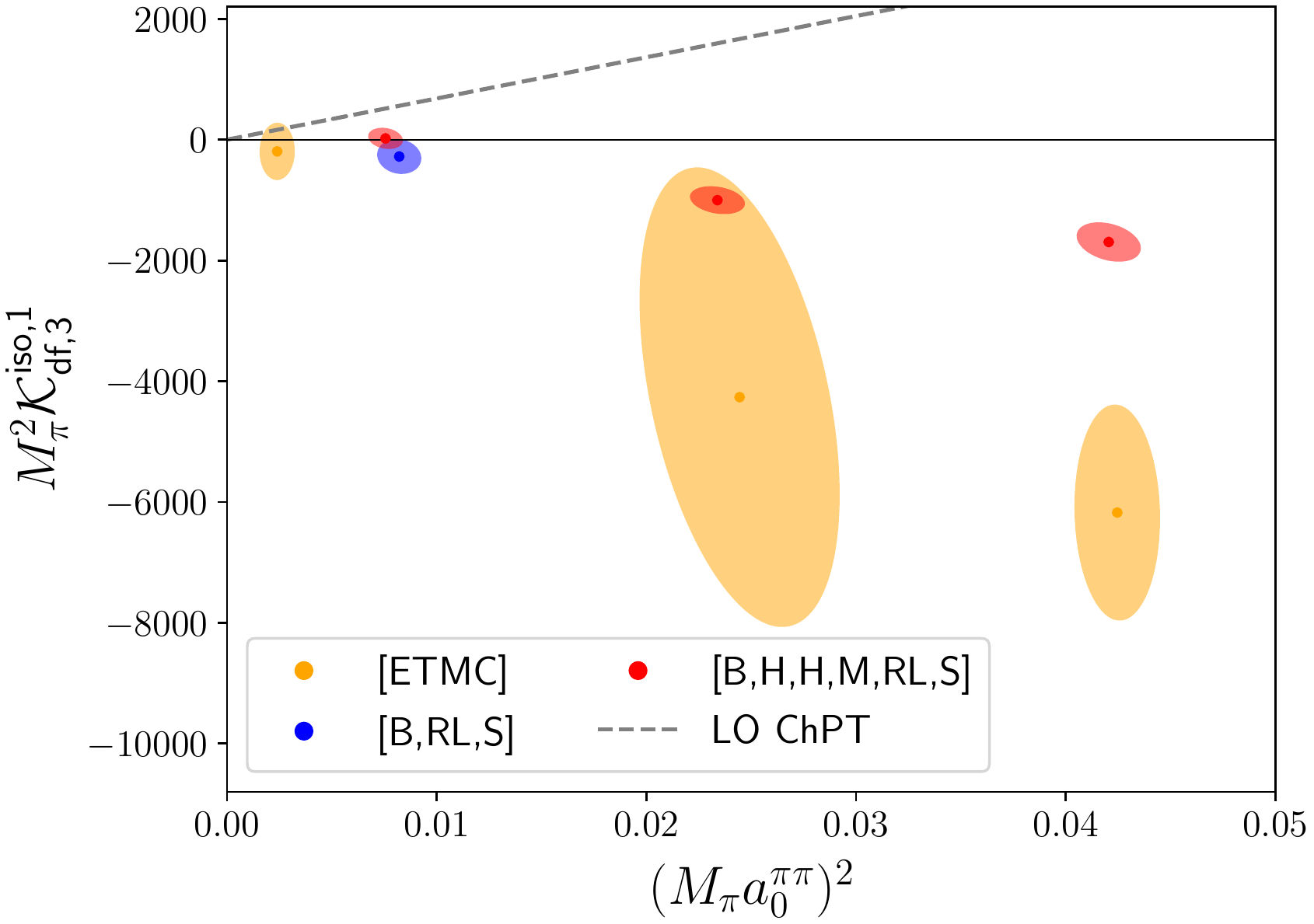}  
  \caption{$\mathcal{K}_{\mathrm{df}, 3}^{\text {iso }, 1}$}
  \label{fig:K1}
\end{subfigure}
\caption{ Summary of results for the first two terms in the expansion of $\mathcal K_\text{df,3}$ for the $3\pi^+$ system, plotted against the $s$-wave isospin-2 pion-pion scattering length. The dashed line shows the LO ChPT prediction of Ref.~\cite{Blanton:2019vdk}. All publications use the RFT formalism, and only $s$-wave interactions are included for the two and three pions. The blue result ([B,RL,S]) is from Ref.~\cite{Blanton:2019vdk}, orange ellipses ([ETMC]) from Ref.~\cite{Fischer:2020jzp}, and the red ones ([B,H,H,M,RL,S]) from Ref.~\cite{Blanton:2021llb}.  Figures taken from Ref.~\cite{Blanton:2021llb}.}
\label{fig:Kiso}
\end{figure}

Let us first consider three pions at maximal isospin ($3\pi^+$). Several works have analyzed the lattice spectra using various different formalisms~\cite{Mai:2018djl,Horz:2019rrn,Blanton:2019vdk,Guo:2020kph,Fischer:2020jzp,Hansen:2020otl,Brett:2021wyd,Blanton:2021llb}. Figure~\ref{fig:Kiso} illustrates a summary of the determinations for $\mathcal{K}_\text{df,3}^{\mathrm{iso}, 0}$  and $\mathcal{K}_\text{df,3}^{\mathrm{iso}, 1}$ using the RFT approach and considering only $s$-wave interactions. The LO ChPT prediction for these quantities, as derived in Ref.~\cite{Blanton:2019vdk}, is also shown. Some points in the plot exhibit statistical significance and there is order-of-magnitude agreement between the different determinations. However, several tensions are also apparent. First, results from different works differ by a few standard deviations, which may be due to discretization effects or the number of flavors used. Second, the LO ChPT prediction does not seem to perform particularly well, especially in $\mathcal{K}_\text{df,3}^{\mathrm{iso}, 1}$, where even the sign disagrees. It would be interesting to see if this can be attributed to higher-order terms in ChPT. Note that while the three-particle amplitude is available at next-to-leading order (NLO) in ChPT~\cite{Bijnens:2021hpq,Bijnens:2022zsq}, the connection to $\mathcal K_\text{df,3}$ is not straightforward and requires further investigation.

\begin{figure}[ht]
\begin{subfigure}{.5\textwidth}
  \centering
  % include first image
  \includegraphics[width=\linewidth]{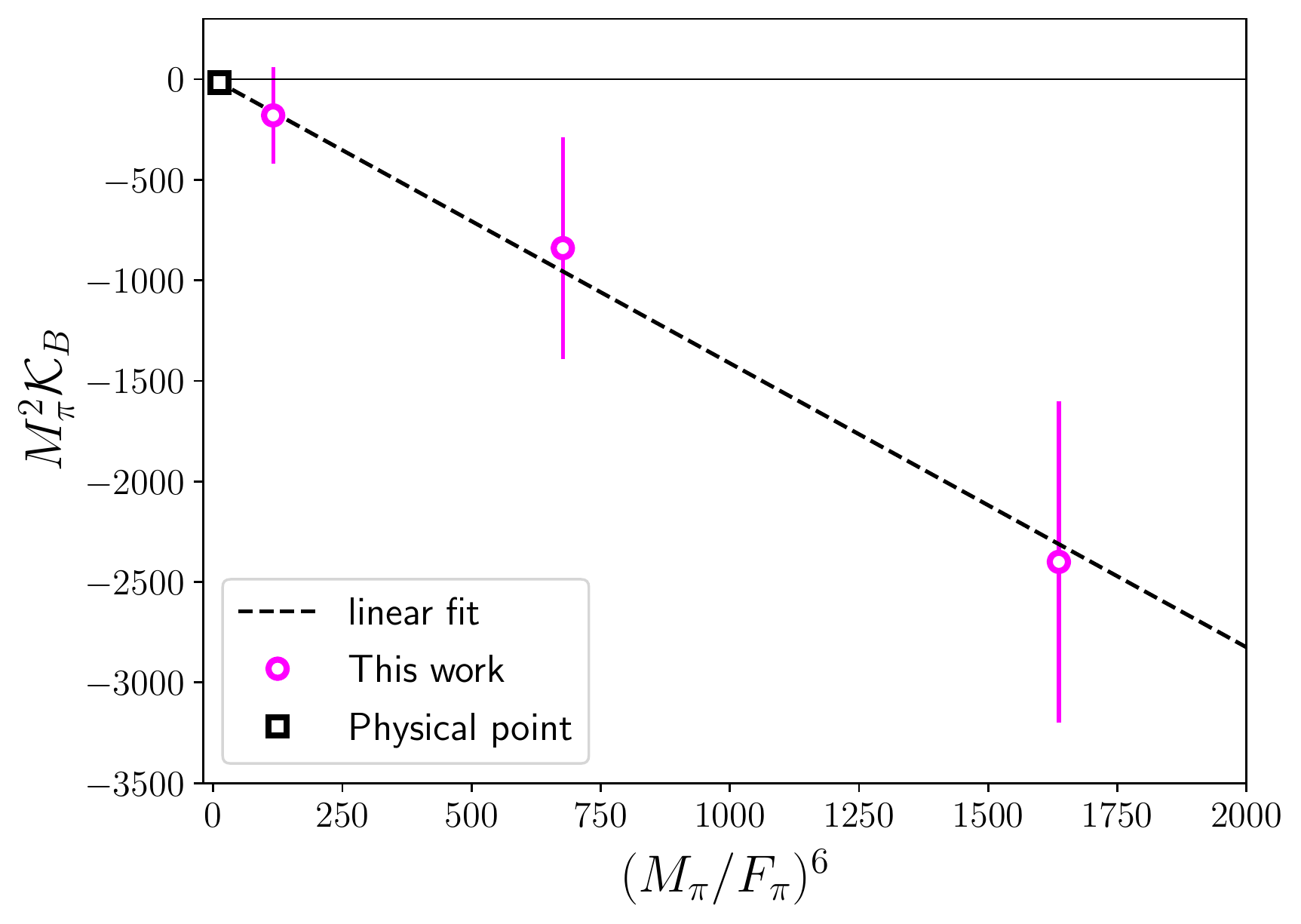}  
  \caption{$\mathcal K_B$ for $3\pi^+$}
  \label{fig:KBpi}
\end{subfigure}
\begin{subfigure}{.5\textwidth}
  \centering
  % include second image
  \includegraphics[width=\linewidth]{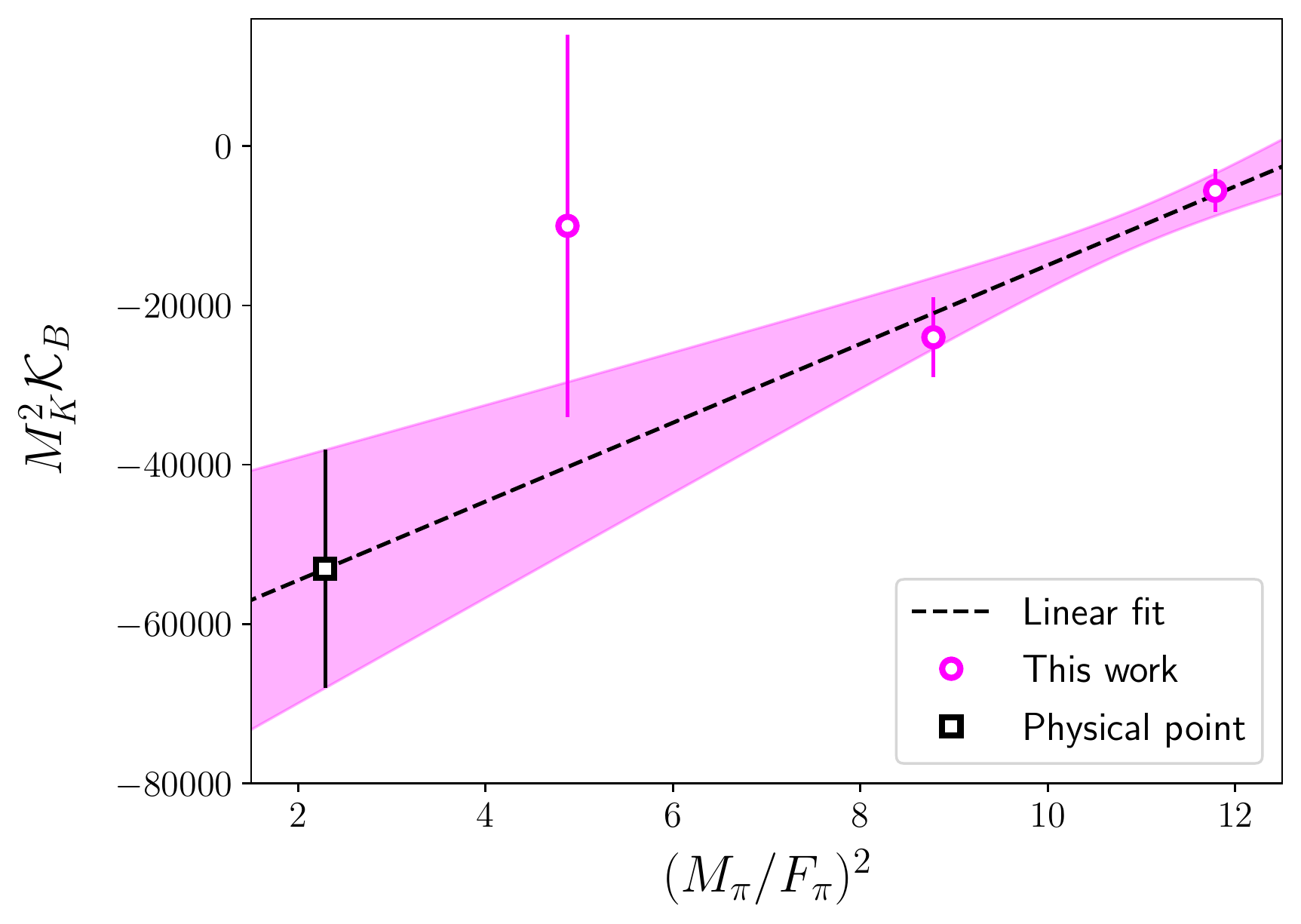}  
  \caption{$\mathcal K_B$ for $3K^+$}
  \label{fig:KBK}
\end{subfigure}
\caption{ Lattice QCD results for $\mathcal K_B$ for three different ensembles in $3\pi^+$ (left) or $3K^+$ (right) systems. The results for three ensembles at different values of the pion mass are shown. An extrapolation to the physical point (black square) is also included, based on the expected chiral behavior (for pions) or a simple linear extrapolation (for kaons). Figures taken from Ref.~\cite{Blanton:2021llb}.}
\label{fig:KB}
\end{figure}

In addition to $3\pi^+$ in an overall $s$ wave, there have been some studies of three positive kaons~\cite{Alexandru:2020xqf,Blanton:2021llb}, as well as three-meson $d$-wave interactions~\cite{Blanton:2021llb}. Specifically, in Ref.~\cite{Blanton:2021llb}, we determined the value of $\mathcal K_B$ for three different ensembles in the $3\pi^+$ and $3K^+$ systems. This is the only term in Eq.~\ref{eq:Kdfdwave} that includes interactions in an overall $d$ wave. The results are presented in Fig.~\ref{fig:KB}, where it can be seen that several of these determinations yield statistically significant values of $\mathcal K_B$. We also include an extrapolation to the physical point. For the case of $3\pi^+$, we used the fact that $\mathcal K_B$ appears first at NLO in ChPT and thus has an expected chiral scaling of $(M_\pi/F_\pi)^6$ up to chiral logarithms. For $3K^+$, a linear extrapolation in $(M_\pi/F_\pi)^2$ was performed. It will be interesting to study these quantities directly at the physical point, especially for kaons, as a rather large result is expected. It would also be useful to derive the NLO ChPT prediction for $\mathcal K_B$ based on Refs.~\cite{Bijnens:2021hpq,Bijnens:2022zsq}.

A class of systems that is very interesting to study are those that contain nondegenerate particles, such as $T_{cc} \to D \pi \pi$. Before considering the complications of resonant channels, it will be useful to focus on three nonidentical weakly-interacting particles, such as $K^+\pi^+\pi^+$ and $K^+K^+\pi^+$. These systems have additional complications compared to $3\pi^+$, such as odd partial waves and different two-particle subchannels: both $\pi^+\pi^+$ and $K^+\pi^+$ subprocesses appear in $K^+\pi^+\pi^+$. The three-particle formalism is similar to that described in Section~\ref{sec:qc3}, except that there is an additional matrix index in the quantization condition labeling the flavor of the spectator particle---see Refs.~\cite{Blanton:2021mih,Blanton:2021eyf} for more detail. Furthermore, as in the case of identical particles, the three-body short-range interaction can be parametrized around the three-particle threshold as:
\begin{equation}
\mathcal{K}_{\mathrm{df}, 3}=\mathcal{K}_0+\mathcal{K}_1 \Delta+\mathcal{K}_B \Delta_3^S+\mathcal{K}_E \tilde{t}_{33}^S+\mathcal{O}\left(\Delta^2\right),
\end{equation}
where $\mathcal{K}_0,\mathcal{K}_1,\mathcal{K}_E$ and $\mathcal{K}_B$ are numerical constants, and $\Delta_3^S$ and $\tilde{t}_{33}$ are kinematic functions---see Eq. 2.27 in Ref.~\cite{Blanton:2021eyf}. Note that the $\mathcal{K}_E$ term is the only one that includes interactions in an overall $p$ wave.

\begin{figure}[ht]
\begin{subfigure}{.5\textwidth}
  \centering
  % include first image
  \includegraphics[width=\linewidth]{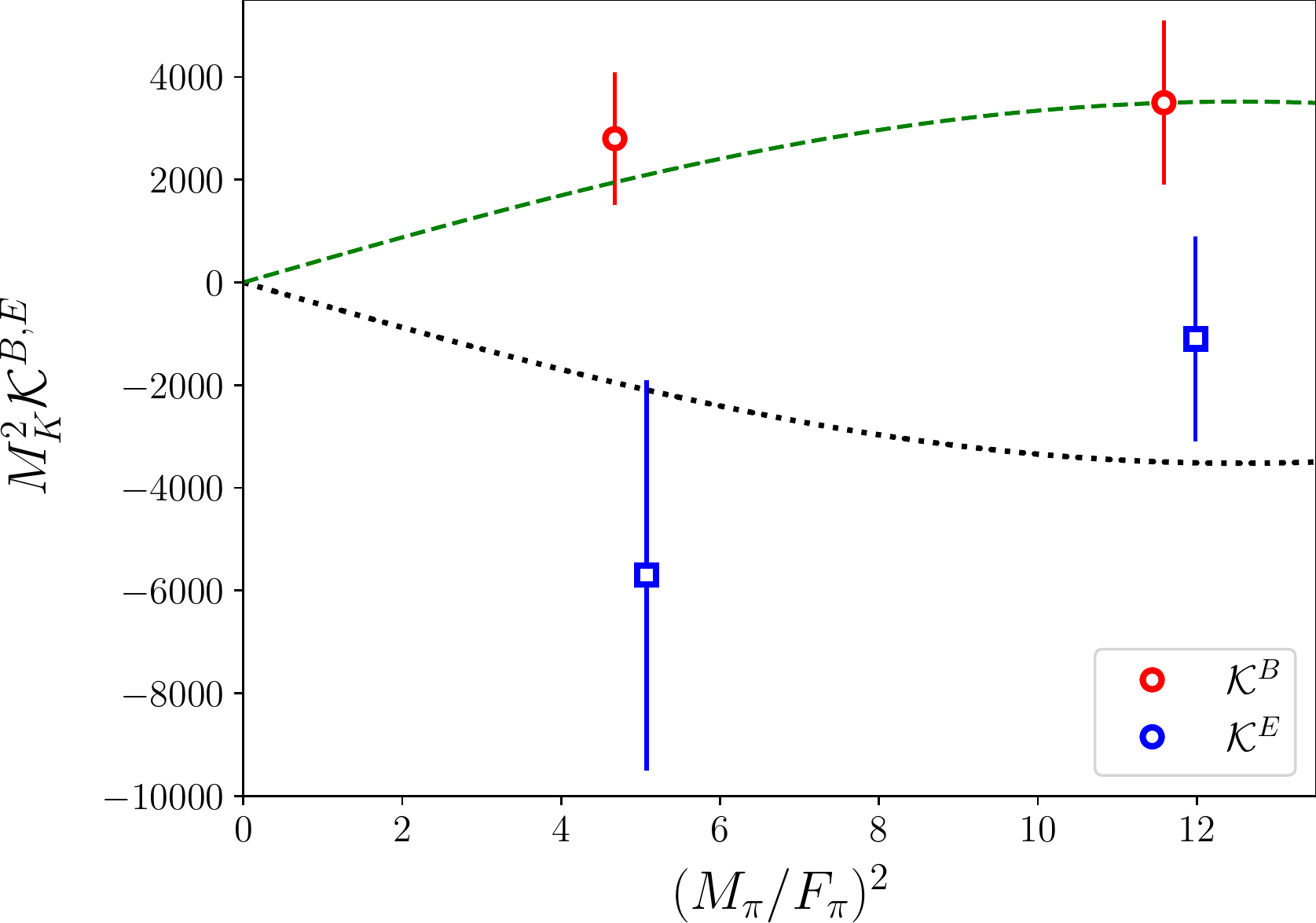}  
  \caption{$K^+K^+\pi^+$}
  \label{fig:KKp}
\end{subfigure}
\begin{subfigure}{.5\textwidth}
  \centering
  % include second image
  \includegraphics[width=\linewidth]{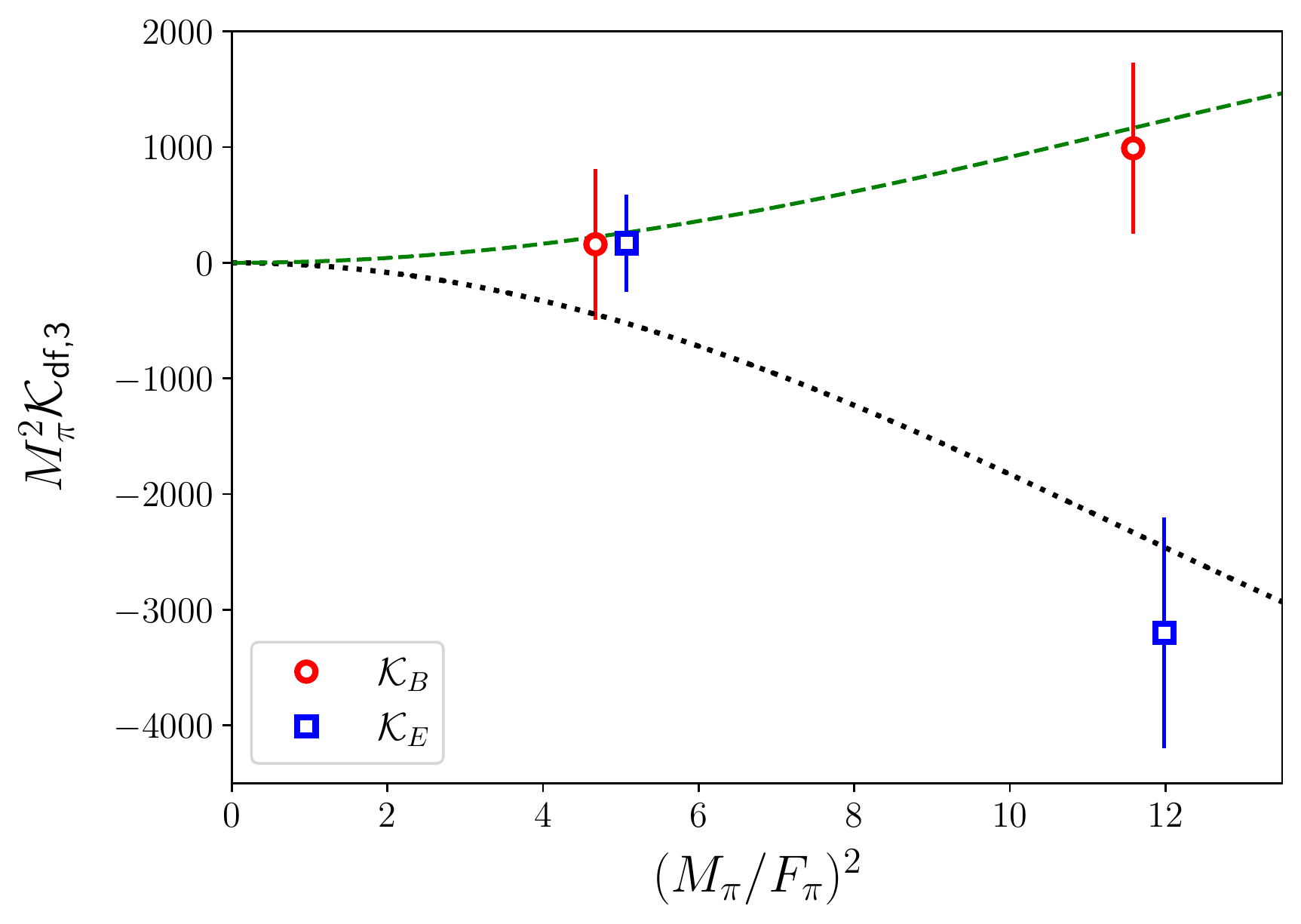}  
  \caption{$K^+\pi^+\pi^+$ }
  \label{fig:ppK}
\end{subfigure}
\caption{ Preliminary results for the $\mathcal K_E$ and $\mathcal K_B$ terms in $\mathcal K_\text{df,3}$ for $K^+K^+\pi^+$ (left) and $K^+\pi^+\pi^+$ (right) systems. The dashed or dotted lines correspond to approximate ChPT expectations (not a fit) based on the leading chiral scaling, as described in Section 4.2 of Ref.~\cite{Blanton:2021eyf}. Figures based on preliminary work presented at this conference by S. Sharpe~\cite{E250inprog}.}
\label{fig:nondeg}
\end{figure}

As an example of ongoing work in this direction, Fig.~\ref{fig:nondeg} displays results for $\mathcal K_B$ and $\mathcal K_E$ for the $K^+K^+\pi^+$ and $K^+\pi^+\pi^+$ systems. By applying the quantization condition derived in Ref.~\cite{Blanton:2021mih}, which was implemented\footnote{See the associated repository~\cite{QC3git}.} in Ref.~\cite{Blanton:2021eyf}, we have analyzed the spectra in two different ensembles with pion masses of approximately 200 MeV and 340 MeV. Our findings indicate that short-range $KK\pi$ interactions are stronger than those for $K\pi\pi$, as anticipated by naive ChPT expectations. In several cases, we also observe results that differ from zero with statistical significance.

\subsection{Three-body resonances}

Weakly-interacting systems are interesting to explore the behavior of three-body systems in general, to benchmark the three-particle formalism, or to compare three-hadron quantities to EFTs. However, a more compelling goal is to explore the properties of resonances with three-body decay modes. In QCD, the first step in that direction has been undertaken in Ref.~\cite{Mai:2021nul} for the $a_0$ resonance, although there is not enough information yet to constrain its properties. An alternative approach has been followed in Ref.~\cite{Garofalo:2022pux}, where a three-particle resonance in a toy model has been investigated to gain further insight into the manifestation of three-particle resonances in finite volume, and how their properties can be extracted using the formalism. 

The toy model of choice is a scalar theory with two complex scalar fields $\varphi_i$, and a term in the Lagrangian that allows for one-to-three transitions:
\begin{equation}
\mathcal L \supset \frac{g}{2} (\varphi_0)^3 \varphi_1^\dagger + \text{h.c.},
\end{equation}
where $g$ is the one-to-three coupling. If the mass of the heavy scalar, $M_1$, is larger than three times that of the light scalar, $M_0$, the heavier particle can be interpreted as a three-particle resonance.

\begin{figure}[h!]
\begin{subfigure}{.5\textwidth}
  \centering
  % include first image
  \includegraphics[width=\linewidth]{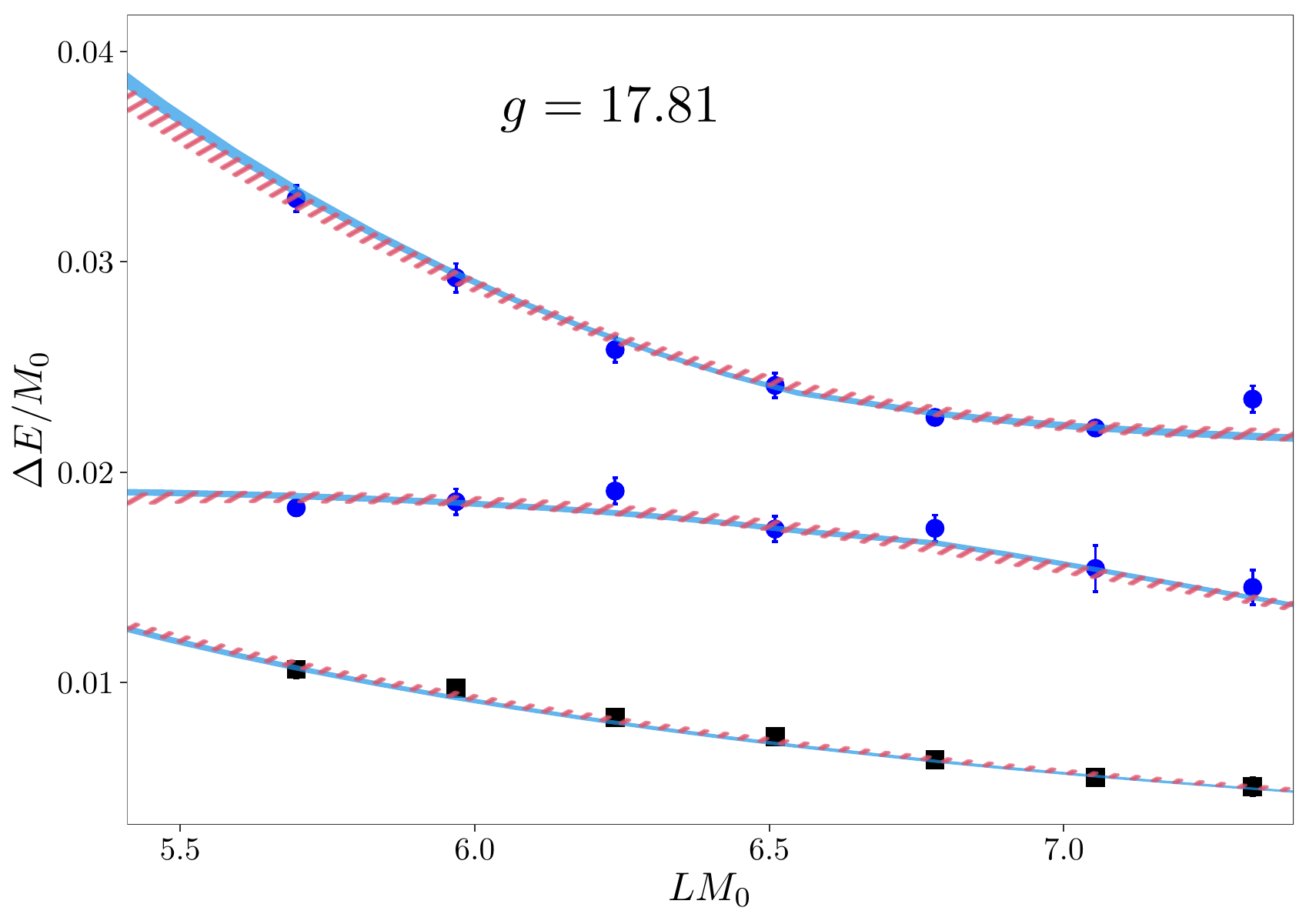}  
  \caption{Spectrum}
  \label{fig:phi4a}
\end{subfigure}
\begin{subfigure}{.5\textwidth}
  \centering
  % include second image
  \includegraphics[width=\linewidth]{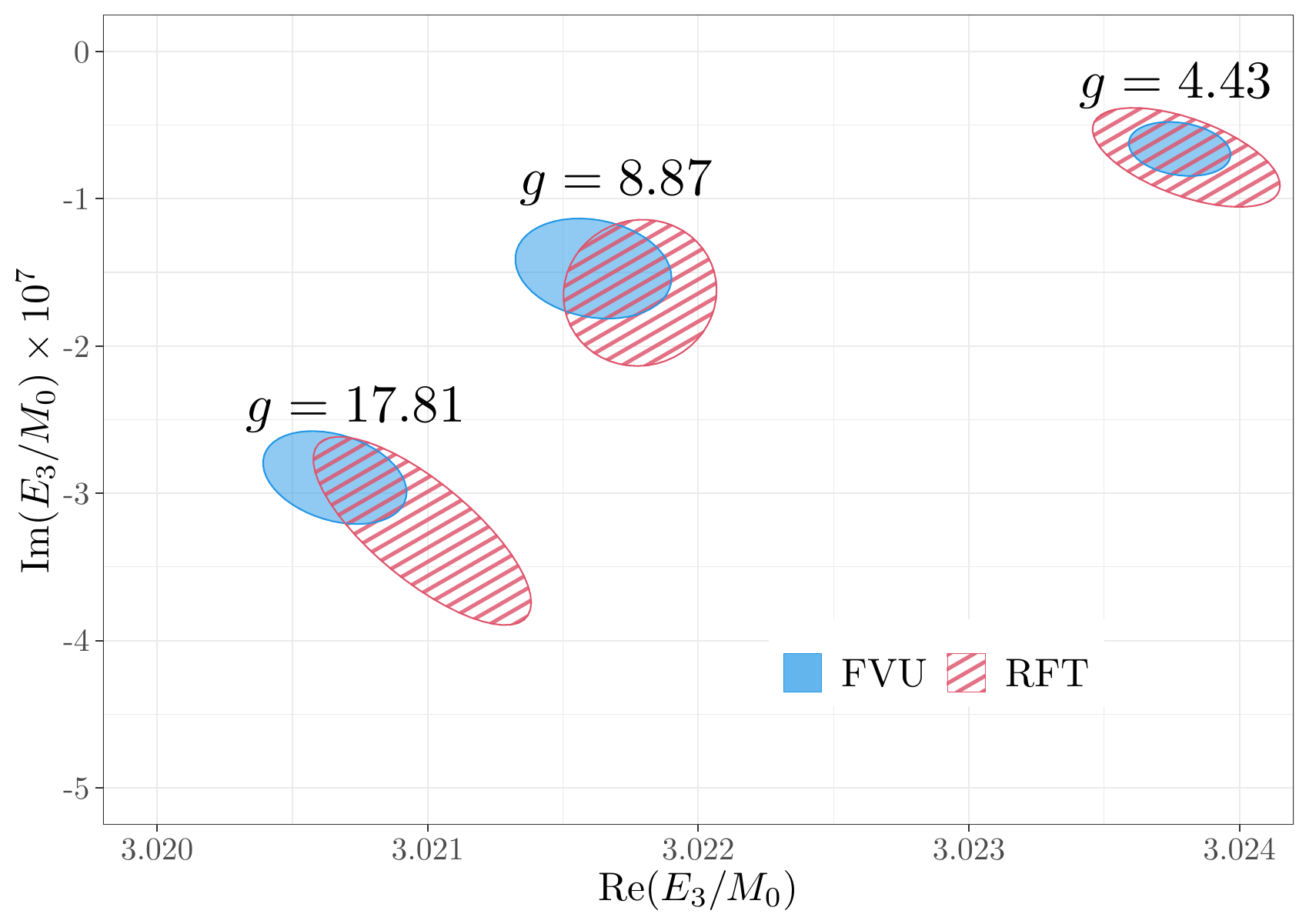}  
  \caption{Pole position}
  \label{fig:phi4b}
\end{subfigure}
\caption{ Left plot: Spectrum for two (black squares) and three (blue circles) particles as a function of the box size, $L$, in units of the mass of the lightest particle, $M_0$ (left). The vertical axes represents the shift with respect to the two- or three-particle threshold. Right plot: Real and imaginary part of the pole position of a resonance in a $\varphi^4$ theory, including simulations at three different values of the $1\to 3$ coupling, $g$. The two different patterns correspond to analysis of the finite-volume spectrum using two alternative methods: FVU for solid blue ellipses, and RFT for red stripped ellipses. Figures taken from Ref.~\cite{Garofalo:2022pux}.}
\label{fig:phi4}
\end{figure}

We perform various lattice simulations at different values of the lattice volume, and at three different sets of parameters in the action. An example is shown in Fig.~\ref{fig:phi4a}, where also the usual imprint of a resonance in finite volume, i.e.~an avoided level crossing, can be seen. In order to analyze the spectrum, we use the quantization condition for identical scalars in both FVU and RFT approaches. We parametrize the three-body K-matrix of the RFT with an explicit pole:
\begin{equation}
\mathcal{K}_{\mathrm{df}, 3}^{\mathrm{iso}}=\frac{c_0^{\prime}}{E_3^2-m_R^{\prime 2}}+c_1^{\prime},
\end{equation}
where $m_R$, $c_0$ and $c_1$ are parameters, and a similar parametrization is used for the analogous quantity in the FVU approach. In order to compute the properties of the resonance, we find the pole position in the complex plane of the three-particle scattering amplitude after solving the corresponding integral equations. More details can be found in Ref.~\cite{Garofalo:2022pux}.

Figure~\ref{fig:phi4b} is the main result of Ref.~\cite{Garofalo:2022pux}. It displays the pole position of the resonance in the complex plane, which is related the mass ($M_R$) and width ($\Gamma$) of the resonance:
\begin{equation}
E_\text{pole} = M_R - i \frac{\Gamma}{2}.
\end{equation}
As can be seen, the width of the resonance increases with the value of $g$. Note that the small value of the width is due to the small phase factor for this transition. In addition, the resonance properties extracted using both formalisms are compatible. Indeed, this is the first demonstration of agreement when analyzing the same lattice spectra with two different three-body finite-volume approaches.

\section{Towards many hadron physics}

There are many processes involving more than three hadrons that are relevant for our understanding of the strong interaction, and the Standard Model in general, for instance, exotic meson resonances, weak decays of D mesons, as well as for binding energies and properties of atomic nuclei. However, the complexity of lattice QCD computations of these quantities is expected to grow with the number of hadrons.

Three ingredients are required for lattice QCD calculations of multi-hadron quantities. First, multi-hadron correlation functions need to be efficiently computed in a situation in which the number of Wick contractions grows factorially with the number of hadrons. Second, large enough statistical samples are required such that observables can be extracted. Third, a theoretical framework that relates finite-volume Euclidean observables to infinite-volume Minkowski multi-hadron quantities. In this section, I will summarize progress on these three fronts.

\begin{figure}[h!]
\centering
\includegraphics[width=0.6\linewidth]{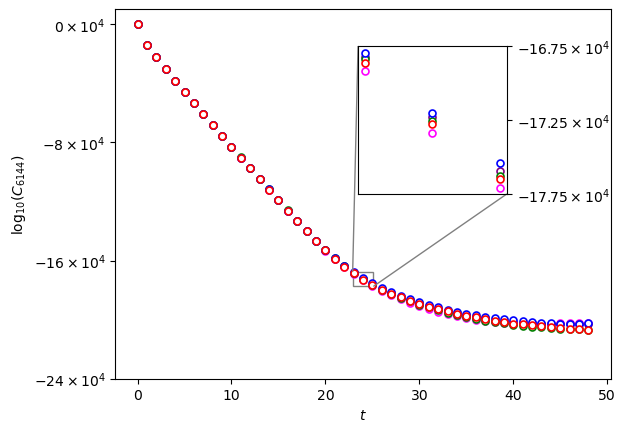}
\caption{Measurement of the 6144-$\pi^+$ correlation function on several configurations. Figure based on work in progress presented at this conference by R. Abbott~\cite{manypi}.}
\label{fig:manypi}
\end{figure}

Several works have addressed the topic of an efficient computation of correlation functions~\cite{Beane:2007qr,Detmold:2008fn,Detmold:2010au,Detmold:2011kw,Detmold:2012pi,Detmold:2014iha}, even up to 72~$\pi^+$~\cite{Detmold:2012pi}. As presented at this conference, we propose and test an algorithm to efficiently compute correlation functions for systems of many $\pi^+$ up to 6144~$\pi^+$. The algorithm scales cubically with the number of pions, and the cost is dominated by a singular-value decomposition. The details will be presented in an upcoming manuscript~\cite{manypi}. An example is given in Fig.~\ref{fig:manypi}, specifically, several measurements in different gauge configurations of the 6144-$\pi^+$ correlation function. As can be seen, the measurements of the correlation function vary over many orders of magnitude, making standard statistical tools completely impractical. Instead, we find that it is possible to obtain information by assuming a log-normal distribution of the correlation function in different configurations~\cite{Endres:2011jm}. Indeed, this is related to ideas that there is information about the theory in the distribution of correlation function, and not only in the mean~\cite{Yunus:2022wuc}.

Regarding the interpretation of many-body Euclidean quantities, there are several complementary approaches that can be pursued. First, a formalism based on quantization conditions, as is the case of two and three particles. This remains so far unexplored. Moreover, alternative approaches to interpret Euclidean correlation functions based on spectral densities have been recently explored~\cite{Hansen:2019idp,Bulava:2019kbi,Bulava:2021fre,Bruno:2020kyl,Garofalo:2021bzl}. Finally, the formulation of EFTs in finite volume can be a useful connection between lattice QCD and infinite-volume quantities~\cite{Detmold:2004qn,Beane:2012ey,Li:2021mob,Li:2019qvh,Severt:2022jtg}. For example, one can use pionless EFT in a finite volume for systems of few nucleons at low momentum, i.e.~$|k| \ll M_\pi$~\cite{Eliyahu:2019nkz,Detmold:2021oro,Sun:2022frr}.

The formulation of pionless EFT in a finite volume is based on the variational principle. The main idea is to have an expressive and trainable set of wave function ans\"atze with periodic boundary conditions, such that the theory is formulated in finite volume. The parameters in the test wave functions are optimized to minimize the energy using, e.g., state-of-the-art machine-learning optimizers. Thus, good approximations for energies and wave functions of the states can be obtained. An example ansatz is having independent Gaussian functions in each spatial direction. 

\begin{figure}[h!]
\centering
\includegraphics[width=0.48\linewidth]{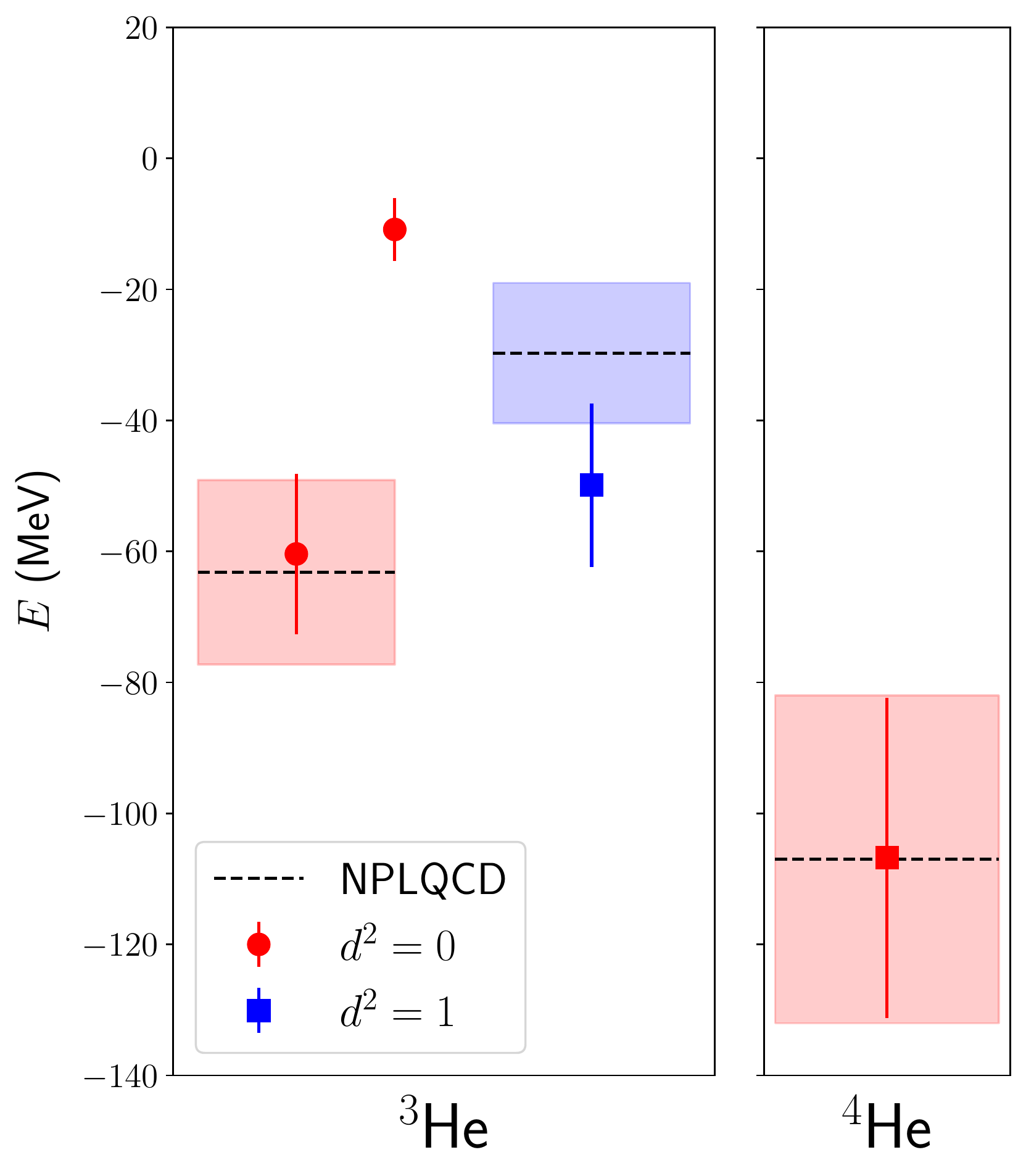}
\caption{Predictions for the energy levels for He-3 and He-4 using pionless EFT in a finite-volume (markers) compared against the lattice QCD data from Ref.~\cite{NPLQCD:2012mex} (dashed black lines and corresponding colored error bands). The effective couplings in two-body sector have been matched to lattice data of the deuteron and dineutron systems, and the three-body coupling has been constrained to the ground state in He-3 in three different volumes---see Ref.~\cite{Sun:2022frr}.}
\label{fig:FVEFT}
\end{figure}

Proof-of-concept results have been shown for up to six-nucleon systems~\cite{Detmold:2021oro,Sun:2022frr}. As presented at this conference, this approach can also be used for excited states and moving frames if the set of test wave functions is large enough and has been adequately optimized. An example is shown in Fig.~\ref{fig:FVEFT} for three- and four-body systems, where the two- and three-body effective couplings have been obtained by fitting to the ground states in the rest frame of the corresponding two- and three-body systems, as described in Ref.~\cite{Sun:2022frr}. The red circles and blue square in this figure correspond to predictions for energies, and in cases where lattice QCD data is available, the predictions agree well with the observed energies.

\section{Summary and Outlook}

Here I have described recent progress towards uncovering the fundamental nature of the hadron spectrum using lattice QCD. In particular, I have discussed how multi-hadron quantities from lattice QCD can be used to compute scattering amplitudes, resonance properties, and effective multi-hadron couplings.

The Lüscher formalism is a well-established tool for studying two-hadron systems. Many two-meson systems have been explored using this formalism, with some computations being carried out directly at the physical point. However, processes involving at least one baryon are more challenging due to the signal-to-noise problem, making it harder to obtain reliable computations of energy levels and requiring large sets of operators. As an example, I have discussed the $\pi N$ scattering amplitude and the $\Delta(1232)$ resonance~\cite{Bulava:2022vpq}.
In the next few years, it is expected that more baryon systems will be explored. It will be important to gain control over the quark mass dependence of scattering amplitudes, as well as the continuum limit. The latter seems to be particularly important, since it has been seen that some scattering amplitudes exhibit significant discretization effects~\cite{Green:2021qol}.

Major developments in the three-particle formalism have enabled the first applications to simple systems, including three-particle resonances~\cite{Mai:2021nul,Garofalo:2022pux}. There is still much potential for QCD applications in this area, and some formal developments are still needed. As a mid-term goal, the Roper resonance captures all expected complications of a three-particle system: (i) it decays into particles with different masses and isospin quantum numbers, pions and a nucleon, (ii) it has both, two- and three-body decay modes, and (iii) it involves hadrons with spin. Nonidentical particles~\cite{Hansen:2020zhy,Blanton:2021mih,Blanton:2020gmf}, and two-to-three transitions~\cite{Briceno:2017tce} have been addressed in isolation, and work on the formalism for spin is ongoing~\cite{3neutrons}. Once all the necessary pieces have been combined, it will be possible to study the analytic structure of the scattering amplitude in the vicinity of the Roper resonance using lattice QCD.

The computations of quantities beyond three hadrons are significantly more complicated, both numerically and theoretically. One useful theoretical tool is the formulation of effective field theories in finite volume. Specifically, pionless EFT can be used as a bridge between few-nucleon quantities computed in finite volume and the infinite-volume properties of mid-sized nuclei such as lithium and beryllium~\cite{Eliyahu:2019nkz,Detmold:2021oro,Sun:2022frr}.

It remains to be seen what the best approach for more than three particles will be, and in fact, the answer may depend on the specific system being studied. However, it is evident that this effort is worth pursuing as it has the potential to provide valuable insights into the properties of exotic hadrons, weak processes with multi-hadron final states, and the emergence of nuclei in QCD.

\section*{Acknowledgements}

I would like to thank the organizers of the Lattice 2022 conference for the opportunity to give this plenary talk. Special thanks to W. Detmold, B. H\"orz, D. Pefkou, A. Rusetsky, S. Sharpe, S. Skinner for feedback on this manuscript, and R. Abbott, S. Sharpe and S. Skinner for help with some plots. This work has been supported in part by the U.S. 
Department of Energy (DOE), Office of Science, Office of Nuclear Physics, under grant Contract Numbers 
DE-SC0011090 and DE-SC0021006.

\bibliographystyle{utphys}
\bibliography{ref.bib}

\providecommand{\href}[2]{#2}\begingroup\raggedright\begin{thebibliography}{100}

\bibitem{Mai:2022eur}
M.~Mai, U.-G. Mei\ss{}ner, and C.~Urbach
  \href{http://arxiv.org/abs/2206.01477}{{\ttfamily arXiv:2206.01477
  [hep-ph]}}.

\bibitem{Wickramaarachchi:2022mhi}
{GlueX} Collaboration, N.~Wickramaarachchi, R.~A. Schumacher, and G.~Kalicy
  \href{http://dx.doi.org/10.1051/epjconf/202227107005}{{\em EPJ Web Conf.}
  {\bfseries 271} (2022) 07005},
  \href{http://arxiv.org/abs/2209.06230}{{\ttfamily arXiv:2209.06230
  [nucl-ex]}}.

\bibitem{Roper:1964zza}
L.~D. Roper \href{http://dx.doi.org/10.1103/PhysRevLett.12.340}{{\em Phys. Rev.
  Lett.} {\bfseries 12} (1964) 340--342}.

\bibitem{LHCb:2021vvq}
{LHCb} Collaboration, R.~Aaij {\em et~al.}
  \href{http://dx.doi.org/10.1038/s41567-022-01614-y}{{\em Nature Phys.}
  {\bfseries 18} no.~7, (2022) 751--754},
  \href{http://arxiv.org/abs/2109.01038}{{\ttfamily arXiv:2109.01038
  [hep-ex]}}.

\bibitem{LHCb:2022xob}
{LHCb} Collaboration \href{http://arxiv.org/abs/2212.02716}{{\ttfamily
  arXiv:2212.02716 [hep-ex]}}.

\bibitem{Ruso:2022qes}
L.~A. Ruso {\em et~al.} \href{http://arxiv.org/abs/2203.09030}{{\ttfamily
  arXiv:2203.09030 [hep-ph]}}.

\bibitem{NA48:1999szy}
{NA48} Collaboration, V.~Fanti {\em et~al.}
  \href{http://dx.doi.org/10.1016/S0370-2693(99)01030-8}{{\em Phys. Lett. B}
  {\bfseries 465} (1999) 335--348},
  \href{http://arxiv.org/abs/hep-ex/9909022}{{\ttfamily arXiv:hep-ex/9909022}}.

\bibitem{KTeV:1999kad}
{KTeV} Collaboration, A.~Alavi-Harati {\em et~al.}
  \href{http://dx.doi.org/10.1103/PhysRevLett.83.22}{{\em Phys. Rev. Lett.}
  {\bfseries 83} (1999) 22--27},
  \href{http://arxiv.org/abs/hep-ex/9905060}{{\ttfamily arXiv:hep-ex/9905060}}.

\bibitem{LHCb:2019hro}
{LHCb} Collaboration, R.~Aaij {\em et~al.}
  \href{http://dx.doi.org/10.1103/PhysRevLett.122.211803}{{\em Phys. Rev.
  Lett.} {\bfseries 122} no.~21, (2019) 211803},
  \href{http://arxiv.org/abs/1903.08726}{{\ttfamily arXiv:1903.08726
  [hep-ex]}}.

\bibitem{Bulava:2022ovd}
J.~Bulava {\em et~al.} in {\em {2022 Snowmass Summer Study}}.
\newblock 3, 2022.
\newblock \href{http://arxiv.org/abs/2203.03230}{{\ttfamily arXiv:2203.03230
  [hep-lat]}}.

\bibitem{Green:2022rjj}
J.~R. Green, A.~D. Hanlon, P.~M. Junnarkar, and H.~Wittig
\newblock 12, 2022.
\newblock \href{http://arxiv.org/abs/2212.09587}{{\ttfamily arXiv:2212.09587
  [hep-lat]}}.

\bibitem{Baeza-Ballesteros:2022bsn}
J.~Baeza-Ballesteros and M.~T. Hansen
\newblock 12, 2022.
\newblock \href{http://arxiv.org/abs/2212.10623}{{\ttfamily arXiv:2212.10623
  [hep-lat]}}.

\bibitem{Jackura:2022xml}
A.~W. Jackura, R.~A. Brice\'no, and M.~T. Hansen
\newblock 12, 2022.
\newblock \href{http://arxiv.org/abs/2212.09951}{{\ttfamily arXiv:2212.09951
  [hep-lat]}}.

\bibitem{Severt:2022eic}
D.~Severt in {\em {39th International Symposium on Lattice Field Theory}}.
\newblock 10, 2022.
\newblock \href{http://arxiv.org/abs/2210.09423}{{\ttfamily arXiv:2210.09423
  [hep-lat]}}.

\bibitem{Wagner:2022bff}
M.~Wagner, C.~Alexandrou, J.~Finkenrath, T.~Leontiou, S.~Meinel, and
  M.~Pflaumer in {\em {39th International Symposium on Lattice Field Theory}}.
\newblock 10, 2022.
\newblock \href{http://arxiv.org/abs/2210.09281}{{\ttfamily arXiv:2210.09281
  [hep-lat]}}.

\bibitem{Bicudo:2022jep}
P.~Bicudo, N.~Cardoso, L.~Mueller, and M.~Wagner in {\em {39th International
  Symposium on Lattice Field Theory}}.
\newblock 10, 2022.
\newblock \href{http://arxiv.org/abs/2210.13284}{{\ttfamily arXiv:2210.13284
  [hep-lat]}}.

\bibitem{Padmanath:2022dxy}
M.~Padmanath, N.~Mathur, and D.~Chakraborty in {\em {39th International
  Symposium on Lattice Field Theory}}.
\newblock 10, 2022.
\newblock \href{http://arxiv.org/abs/2210.13026}{{\ttfamily arXiv:2210.13026
  [hep-lat]}}.

\bibitem{Riederer:2022apu}
B.~Riederer and A.~Maas \href{http://dx.doi.org/10.22323/1.414.0878}{{\em PoS}
  {\bfseries ICHEP2022} (2022) 878},
  \href{http://arxiv.org/abs/2210.17211}{{\ttfamily arXiv:2210.17211
  [hep-lat]}}.

\bibitem{Hoffmann:2022jdx}
J.~Hoffmann, A.~Zimermmane-Santos, and M.~Wagner in {\em {39th International
  Symposium on Lattice Field Theory}}.
\newblock 11, 2022.
\newblock \href{http://arxiv.org/abs/2211.15765}{{\ttfamily arXiv:2211.15765
  [hep-lat]}}.

\bibitem{Aoki:2022xxq}
S.~Aoki and T.~Aoki in {\em {39th International Symposium on Lattice Field
  Theory}}.
\newblock 11, 2022.
\newblock \href{http://arxiv.org/abs/2212.00202}{{\ttfamily arXiv:2212.00202
  [hep-lat]}}.

\bibitem{Sadl:2022bnq}
M.~Sadl, S.~Collins, M.~Padmanath, and S.~Prelovsek in {\em {39th International
  Symposium on Lattice Field Theory}}.
\newblock 12, 2022.
\newblock \href{http://arxiv.org/abs/2212.04835}{{\ttfamily arXiv:2212.04835
  [hep-lat]}}.

\bibitem{Bulava:2023mjc}
J.~Bulava
\newblock 1, 2023.
\newblock \href{http://arxiv.org/abs/2301.04072}{{\ttfamily arXiv:2301.04072
  [hep-lat]}}.

\bibitem{Raposo:2023nex}
A.~B.~a. Raposo and M.~T. Hansen
\newblock 1, 2023.
\newblock \href{http://arxiv.org/abs/2301.03981}{{\ttfamily arXiv:2301.03981
  [hep-lat]}}.

\bibitem{Luscher:1986pf}
M.~Luscher \href{http://dx.doi.org/10.1007/BF01211097}{{\em Commun. Math.
  Phys.} {\bfseries 105} (1986) 153--188}.

\bibitem{Luscher:1991cf}
M.~Luscher \href{http://dx.doi.org/10.1016/0550-3213(91)90584-K}{{\em Nucl.
  Phys. B} {\bfseries 364} (1991) 237--251}.

\bibitem{Ishii:2006ec}
N.~Ishii, S.~Aoki, and T.~Hatsuda
  \href{http://dx.doi.org/10.1103/PhysRevLett.99.022001}{{\em Phys. Rev. Lett.}
  {\bfseries 99} (2007) 022001},
  \href{http://arxiv.org/abs/nucl-th/0611096}{{\ttfamily
  arXiv:nucl-th/0611096}}.

\bibitem{Aoki:2009ji}
S.~Aoki, T.~Hatsuda, and N.~Ishii
  \href{http://dx.doi.org/10.1143/PTP.123.89}{{\em Prog. Theor. Phys.}
  {\bfseries 123} (2010) 89--128},
  \href{http://arxiv.org/abs/0909.5585}{{\ttfamily arXiv:0909.5585 [hep-lat]}}.

\bibitem{Aoki:2012tk}
{HAL QCD} Collaboration, S.~Aoki, T.~Doi, T.~Hatsuda, Y.~Ikeda, T.~Inoue,
  N.~Ishii, K.~Murano, H.~Nemura, and K.~Sasaki
  \href{http://dx.doi.org/10.1093/ptep/pts010}{{\em PTEP} {\bfseries 2012}
  (2012) 01A105}, \href{http://arxiv.org/abs/1206.5088}{{\ttfamily
  arXiv:1206.5088 [hep-lat]}}.

\bibitem{Rummukainen:1995vs}
K.~Rummukainen and S.~A. Gottlieb
  \href{http://dx.doi.org/10.1016/0550-3213(95)00313-H}{{\em Nucl. Phys.}
  {\bfseries B450} (1995) 397--436},
\href{http://arxiv.org/abs/hep-lat/9503028}{{\ttfamily arXiv:hep-lat/9503028
  [hep-lat]}}.
%%CITATION = HEP-LAT/9503028;%%.

\bibitem{Kim:2005gf}
C.~h. Kim, C.~T. Sachrajda, and S.~R. Sharpe
  \href{http://dx.doi.org/10.1016/j.nuclphysb.2005.08.029}{{\em Nucl. Phys.}
  {\bfseries B727} (2005) 218--243},
\href{http://arxiv.org/abs/hep-lat/0507006}{{\ttfamily arXiv:hep-lat/0507006
  [hep-lat]}}.
%%CITATION = HEP-LAT/0507006;%%.

\bibitem{He:2005ey}
S.~He, X.~Feng, and C.~Liu
  \href{http://dx.doi.org/10.1088/1126-6708/2005/07/011}{{\em JHEP} {\bfseries
  07} (2005) 011},
\href{http://arxiv.org/abs/hep-lat/0504019}{{\ttfamily arXiv:hep-lat/0504019
  [hep-lat]}}.
%%CITATION = HEP-LAT/0504019;%%.

\bibitem{Bernard:2010fp}
V.~Bernard, M.~Lage, U.~G. Mei{$\ss$}ner, and A.~Rusetsky
  \href{http://dx.doi.org/10.1007/JHEP01(2011)019}{{\em JHEP} {\bfseries 01}
  (2011) 019},
\href{http://arxiv.org/abs/1010.6018}{{\ttfamily arXiv:1010.6018 [hep-lat]}}.
%%CITATION = ARXIV:1010.6018;%%.

\bibitem{Hansen:2012tf}
M.~T. Hansen and S.~R. Sharpe
  \href{http://dx.doi.org/10.1103/PhysRevD.86.016007}{{\em Phys. Rev. D}
  {\bfseries 86} (2012) 016007},
  \href{http://arxiv.org/abs/1204.0826}{{\ttfamily arXiv:1204.0826 [hep-lat]}}.

\bibitem{Briceno:2012yi}
R.~A. Brice\~no and Z.~Davoudi
  \href{http://dx.doi.org/10.1103/PhysRevD.88.094507}{{\em Phys. Rev.}
  {\bfseries D88} no.~9, (2013) 094507},
\href{http://arxiv.org/abs/1204.1110}{{\ttfamily arXiv:1204.1110 [hep-lat]}}.
%%CITATION = ARXIV:1204.1110;%%.

\bibitem{Briceno:2014oea}
R.~A. Brice\~no \href{http://dx.doi.org/10.1103/PhysRevD.89.074507}{{\em Phys.
  Rev.} {\bfseries D89} no.~7, (2014) 074507},
\href{http://arxiv.org/abs/1401.3312}{{\ttfamily arXiv:1401.3312 [hep-lat]}}.
%%CITATION = ARXIV:1401.3312;%%.

\bibitem{Romero-Lopez:2018zyy}
F.~Romero-L\'opez, A.~Rusetsky, and C.~Urbach
  \href{http://dx.doi.org/10.1103/PhysRevD.98.014503}{{\em Phys. Rev.}
  {\bfseries D98} no.~1, (2018) 014503},
\href{http://arxiv.org/abs/1802.03458}{{\ttfamily arXiv:1802.03458 [hep-lat]}}.
%%CITATION = ARXIV:1802.03458;%%.

\bibitem{Woss:2020cmp}
A.~J. Woss, D.~J. Wilson, and J.~J. Dudek
\href{http://arxiv.org/abs/2001.08474}{{\ttfamily arXiv:2001.08474 [hep-lat]}}.
%%CITATION = ARXIV:2001.08474;%%.

\bibitem{Grabowska:2021xkp}
D.~M. Grabowska and M.~T. Hansen
  \href{http://dx.doi.org/10.1007/JHEP09(2022)232}{{\em JHEP} {\bfseries 09}
  (2022) 232}, \href{http://arxiv.org/abs/2110.06878}{{\ttfamily
  arXiv:2110.06878 [hep-lat]}}.

\bibitem{Blanton:2019vdk}
T.~D. Blanton, F.~Romero-L\'opez, and S.~R. Sharpe
  \href{http://dx.doi.org/10.1103/PhysRevLett.124.032001}{{\em Phys. Rev.
  Lett.} {\bfseries 124} no.~3, (2020) 032001},
\href{http://arxiv.org/abs/1909.02973}{{\ttfamily arXiv:1909.02973 [hep-lat]}}.
%%CITATION = ARXIV:1909.02973;%%.

\bibitem{Colangelo:2001df}
G.~Colangelo, J.~Gasser, and H.~Leutwyler
  \href{http://dx.doi.org/10.1016/S0550-3213(01)00147-X}{{\em Nucl. Phys. B}
  {\bfseries 603} (2001) 125--179},
  \href{http://arxiv.org/abs/hep-ph/0103088}{{\ttfamily arXiv:hep-ph/0103088}}.

\bibitem{Kaminski:2006qe}
R.~Kaminski, J.~R. Pelaez, and F.~J. Yndurain
  \href{http://dx.doi.org/10.1103/PhysRevD.77.054015}{{\em Phys. Rev. D}
  {\bfseries 77} (2008) 054015},
  \href{http://arxiv.org/abs/0710.1150}{{\ttfamily arXiv:0710.1150 [hep-ph]}}.

\bibitem{Fischer:2020jzp}
M.~Fischer, B.~Kostrzewa, L.~Liu, F.~Romero-L\'opez, M.~Ueding, and C.~Urbach
  \href{http://dx.doi.org/10.1140/epjc/s10052-021-09206-5}{{\em Eur. Phys. J.
  C} {\bfseries 81} no.~5, (2021) 436},
  \href{http://arxiv.org/abs/2008.03035}{{\ttfamily arXiv:2008.03035
  [hep-lat]}}.

\bibitem{E250inprog}
Z.~T. Draper, A.~D. Hanlon, B.~H\"orz, C.~Morningstar, F.~Romero-L\'opez, and
  S.~R. Sharpe
\newblock 2022.
\newblock \href{http://arxiv.org/abs/in progress}{{\ttfamily in progress}}.

\bibitem{Fischer:2020yvw}
{Extended Twisted Mass, ETM} Collaboration, M.~Fischer, B.~Kostrzewa, M.~Mai,
  M.~Petschlies, F.~Pittler, M.~Ueding, C.~Urbach, and M.~Werner
  \href{http://dx.doi.org/10.1016/j.physletb.2021.136449}{{\em Phys. Lett. B}
  {\bfseries 819} (2021) 136449},
  \href{http://arxiv.org/abs/2006.13805}{{\ttfamily arXiv:2006.13805
  [hep-lat]}}.

\bibitem{RBC:2021acc}
{RBC, UKQCD} Collaboration, T.~Blum {\em et~al.}
  \href{http://dx.doi.org/10.1103/PhysRevD.104.114506}{{\em Phys. Rev. D}
  {\bfseries 104} no.~11, (2021) 114506},
  \href{http://arxiv.org/abs/2103.15131}{{\ttfamily arXiv:2103.15131
  [hep-lat]}}.

\bibitem{Bali:2015gji}
{RQCD} Collaboration, G.~S. Bali, S.~Collins, A.~Cox, G.~Donald, M.~G\"ockeler,
  C.~B. Lang, and A.~Sch\"afer
  \href{http://dx.doi.org/10.1103/PhysRevD.93.054509}{{\em Phys. Rev. D}
  {\bfseries 93} no.~5, (2016) 054509},
  \href{http://arxiv.org/abs/1512.08678}{{\ttfamily arXiv:1512.08678
  [hep-lat]}}.

\bibitem{Briceno:2017max}
R.~A. Briceno, J.~J. Dudek, and R.~D. Young
  \href{http://dx.doi.org/10.1103/RevModPhys.90.025001}{{\em Rev. Mod. Phys.}
  {\bfseries 90} no.~2, (2018) 025001},
  \href{http://arxiv.org/abs/1706.06223}{{\ttfamily arXiv:1706.06223
  [hep-lat]}}.

\bibitem{Horz:2022glt}
B.~H\"orz \href{http://dx.doi.org/10.22323/1.396.0006}{{\em PoS} {\bfseries
  LATTICE2021} (2022) 006}.

\bibitem{Orginos:2015aya}
K.~Orginos, A.~Parreno, M.~J. Savage, S.~R. Beane, E.~Chang, and W.~Detmold
  \href{http://dx.doi.org/10.1103/PhysRevD.92.114512}{{\em Phys. Rev. D}
  {\bfseries 92} no.~11, (2015) 114512},
  \href{http://arxiv.org/abs/1508.07583}{{\ttfamily arXiv:1508.07583
  [hep-lat]}}. [Erratum: Phys.Rev.D 102, 039903 (2020)].

\bibitem{NPLQCD:2020lxg}
{NPLQCD} Collaboration, M.~Illa {\em et~al.}
  \href{http://dx.doi.org/10.1103/PhysRevD.103.054508}{{\em Phys. Rev. D}
  {\bfseries 103} no.~5, (2021) 054508},
  \href{http://arxiv.org/abs/2009.12357}{{\ttfamily arXiv:2009.12357
  [hep-lat]}}.

\bibitem{Francis:2018qch}
A.~Francis, J.~R. Green, P.~M. Junnarkar, C.~Miao, T.~D. Rae, and H.~Wittig
  \href{http://dx.doi.org/10.1103/PhysRevD.99.074505}{{\em Phys. Rev. D}
  {\bfseries 99} no.~7, (2019) 074505},
  \href{http://arxiv.org/abs/1805.03966}{{\ttfamily arXiv:1805.03966
  [hep-lat]}}.

\bibitem{Green:2021qol}
J.~R. Green, A.~D. Hanlon, P.~M. Junnarkar, and H.~Wittig
  \href{http://arxiv.org/abs/2103.01054}{{\ttfamily arXiv:2103.01054
  [hep-lat]}}.

\bibitem{Horz:2020zvv}
B.~H\"orz {\em et~al.}
  \href{http://dx.doi.org/10.1103/PhysRevC.103.014003}{{\em Phys. Rev. C}
  {\bfseries 103} no.~1, (2021) 014003},
  \href{http://arxiv.org/abs/2009.11825}{{\ttfamily arXiv:2009.11825
  [hep-lat]}}.

\bibitem{Amarasinghe:2021lqa}
S.~Amarasinghe, R.~Baghdadi, Z.~Davoudi, W.~Detmold, M.~Illa, A.~Parreno, A.~V.
  Pochinsky, P.~E. Shanahan, and M.~L. Wagman
  \href{http://arxiv.org/abs/2108.10835}{{\ttfamily arXiv:2108.10835
  [hep-lat]}}.

\bibitem{Fukugita:1994ve}
M.~Fukugita, Y.~Kuramashi, M.~Okawa, H.~Mino, and A.~Ukawa
  \href{http://dx.doi.org/10.1103/PhysRevD.52.3003}{{\em Phys. Rev. D}
  {\bfseries 52} (1995) 3003--3023},
  \href{http://arxiv.org/abs/hep-lat/9501024}{{\ttfamily
  arXiv:hep-lat/9501024}}.

\bibitem{Lang:2012db}
C.~B. Lang and V.~Verduci
  \href{http://dx.doi.org/10.1103/PhysRevD.87.054502}{{\em Phys. Rev. D}
  {\bfseries 87} no.~5, (2013) 054502},
  \href{http://arxiv.org/abs/1212.5055}{{\ttfamily arXiv:1212.5055 [hep-lat]}}.

\bibitem{Lang:2016hnn}
C.~B. Lang, L.~Leskovec, M.~Padmanath, and S.~Prelovsek
  \href{http://dx.doi.org/10.1103/PhysRevD.95.014510}{{\em Phys. Rev. D}
  {\bfseries 95} no.~1, (2017) 014510},
  \href{http://arxiv.org/abs/1610.01422}{{\ttfamily arXiv:1610.01422
  [hep-lat]}}.

\bibitem{Andersen:2017una}
C.~W. Andersen, J.~Bulava, B.~H\"{o}rz, and C.~Morningstar
\href{http://arxiv.org/abs/1710.01557}{{\ttfamily arXiv:1710.01557 [hep-lat]}}.
%%CITATION = ARXIV:1710.01557;%%.

\bibitem{Alexandrou:2017mpi}
C.~Alexandrou, L.~Leskovec, S.~Meinel, J.~Negele, S.~Paul, M.~Petschlies,
  A.~Pochinsky, G.~Rendon, and S.~Syritsyn
  \href{http://dx.doi.org/10.1103/PhysRevD.96.034525}{{\em Phys. Rev. D}
  {\bfseries 96} no.~3, (2017) 034525},
  \href{http://arxiv.org/abs/1704.05439}{{\ttfamily arXiv:1704.05439
  [hep-lat]}}.

\bibitem{Verduci:2014btc}
V.~Verduci.
\newblock PhD thesis, Graz U., 2014.

\bibitem{Mohler:2012nh}
D.~Mohler \href{http://dx.doi.org/10.22323/1.164.0003}{{\em PoS} {\bfseries
  LATTICE2012} (2012) 003}, \href{http://arxiv.org/abs/1211.6163}{{\ttfamily
  arXiv:1211.6163 [hep-lat]}}.

\bibitem{Pittler:2021bqw}
F.~Pittler, C.~Alexandrou, K.~Hadjiannakou, G.~Koutsou, S.~Paul, M.~Petschlies,
  and A.~Todaro \href{http://dx.doi.org/10.22323/1.396.0226}{{\em PoS}
  {\bfseries LATTICE2021} (2022) 226},
  \href{http://arxiv.org/abs/2112.04146}{{\ttfamily arXiv:2112.04146
  [hep-lat]}}.

\bibitem{Detmold:2015qwf}
W.~Detmold and A.~Nicholson
  \href{http://dx.doi.org/10.1103/PhysRevD.93.114511}{{\em Phys. Rev. D}
  {\bfseries 93} no.~11, (2016) 114511},
  \href{http://arxiv.org/abs/1511.02275}{{\ttfamily arXiv:1511.02275
  [hep-lat]}}.

\bibitem{Torok:2009dg}
A.~Torok, S.~R. Beane, W.~Detmold, T.~C. Luu, K.~Orginos, A.~Parreno, M.~J.
  Savage, and A.~Walker-Loud
  \href{http://dx.doi.org/10.1103/PhysRevD.81.074506}{{\em Phys. Rev. D}
  {\bfseries 81} (2010) 074506},
  \href{http://arxiv.org/abs/0907.1913}{{\ttfamily arXiv:0907.1913 [hep-lat]}}.

\bibitem{Bulava:2022vpq}
J.~Bulava, A.~D. Hanlon, B.~H\"orz, C.~Morningstar, A.~Nicholson,
  F.~Romero-L\'opez, S.~Skinner, P.~Vranas, and A.~Walker-Loud
  \href{http://arxiv.org/abs/2208.03867}{{\ttfamily arXiv:2208.03867
  [hep-lat]}}.

\bibitem{Peardon:2009gh}
{Hadron Spectrum} Collaboration, M.~Peardon, J.~Bulava, J.~Foley,
  C.~Morningstar, J.~Dudek, R.~G. Edwards, B.~Joo, H.-W. Lin, D.~G. Richards,
  and K.~J. Juge \href{http://dx.doi.org/10.1103/PhysRevD.80.054506}{{\em Phys.
  Rev. D} {\bfseries 80} (2009) 054506},
  \href{http://arxiv.org/abs/0905.2160}{{\ttfamily arXiv:0905.2160 [hep-lat]}}.

\bibitem{Morningstar:2011ka}
C.~Morningstar, J.~Bulava, J.~Foley, K.~J. Juge, D.~Lenkner, M.~Peardon, and
  C.~H. Wong \href{http://dx.doi.org/10.1103/PhysRevD.83.114505}{{\em Phys.
  Rev. D} {\bfseries 83} (2011) 114505},
  \href{http://arxiv.org/abs/1104.3870}{{\ttfamily arXiv:1104.3870 [hep-lat]}}.

\bibitem{Silvi:2021}
G.~Silvi, S.~Paul, C.~Alexandrou, S.~Krieg, L.~Leskovec, S.~Meinel, J.~Negele,
  M.~Petschlies, A.~Pochinsky, G.~Rendon, S.~Syritsyn, and A.~Todaro
  \href{http://dx.doi.org/10.1103/physrevd.103.094508}{{\em Physical Review D}
  {\bfseries 103} (May, 2021) }.
  \url{https://doi.org/10.1103%2Fphysrevd.103.094508}.

\bibitem{pdg:2022}
P.~A. Zyla {\em et~al.} \href{http://dx.doi.org/10.1093/ptep/ptaa104}{{\em
  Progress of Theoretical and Experimental Physics} {\bfseries 2020} (08, 2020)
  },
  \href{http://arxiv.org/abs/https://academic.oup.com/ptep/article-pdf/2020/8/083C01/34673722/ptaa104.pdf}{{\ttfamily
  https://academic.oup.com/ptep/article-pdf/2020/8/083C01/34673722/ptaa104.pdf}}.
  \url{https://doi.org/10.1093/ptep/ptaa104}. 083C01.

\bibitem{Hemmert:1994ky}
T.~R. Hemmert, B.~R. Holstein, and N.~C. Mukhopadhyay
  \href{http://dx.doi.org/10.1103/PhysRevD.51.158}{{\em Phys. Rev. D}
  {\bfseries 51} (1995) 158--167},
  \href{http://arxiv.org/abs/hep-ph/9409323}{{\ttfamily arXiv:hep-ph/9409323}}.

\bibitem{Hoferichter:2015tha}
M.~Hoferichter, J.~Ruiz~de Elvira, B.~Kubis, and U.-G. Mei\ss{}ner
  \href{http://dx.doi.org/10.1103/PhysRevLett.115.192301}{{\em Phys. Rev.
  Lett.} {\bfseries 115} no.~19, (2015) 192301},
  \href{http://arxiv.org/abs/1507.07552}{{\ttfamily arXiv:1507.07552
  [nucl-th]}}.

\bibitem{Lellouch:2000pv}
L.~Lellouch and M.~L\"{u}scher
  \href{http://dx.doi.org/10.1007/s002200100410}{{\em Commun. Math. Phys.}
  {\bfseries 219} (2001) 31--44},
  \href{http://arxiv.org/abs/hep-lat/0003023}{{\ttfamily
  arXiv:hep-lat/0003023}}.

\bibitem{Christ:2005gi}
N.~H. Christ, C.~Kim, and T.~Yamazaki
  \href{http://dx.doi.org/10.1103/PhysRevD.72.114506}{{\em Phys. Rev. D}
  {\bfseries 72} (2005) 114506},
  \href{http://arxiv.org/abs/hep-lat/0507009}{{\ttfamily
  arXiv:hep-lat/0507009}}.

\bibitem{Briceno:2014uqa}
R.~A. Brice\~no, M.~T. Hansen, and A.~Walker-Loud
  \href{http://dx.doi.org/10.1103/PhysRevD.91.034501}{{\em Phys. Rev. D}
  {\bfseries 91} no.~3, (2015) 034501},
  \href{http://arxiv.org/abs/1406.5965}{{\ttfamily arXiv:1406.5965 [hep-lat]}}.

\bibitem{Briceno:2015csa}
R.~A. Brice\~no and M.~T. Hansen
  \href{http://dx.doi.org/10.1103/PhysRevD.92.074509}{{\em Phys. Rev. D}
  {\bfseries 92} no.~7, (2015) 074509},
  \href{http://arxiv.org/abs/1502.04314}{{\ttfamily arXiv:1502.04314
  [hep-lat]}}.

\bibitem{Briceno:2021xlc}
R.~A. Brice\~no, J.~J. Dudek, and L.~Leskovec
  \href{http://dx.doi.org/10.1103/PhysRevD.104.054509}{{\em Phys. Rev. D}
  {\bfseries 104} no.~5, (2021) 054509},
  \href{http://arxiv.org/abs/2105.02017}{{\ttfamily arXiv:2105.02017
  [hep-lat]}}.

\bibitem{Feng:2014gba}
X.~Feng, S.~Aoki, S.~Hashimoto, and T.~Kaneko
  \href{http://dx.doi.org/10.1103/PhysRevD.91.054504}{{\em Phys. Rev. D}
  {\bfseries 91} no.~5, (2015) 054504},
  \href{http://arxiv.org/abs/1412.6319}{{\ttfamily arXiv:1412.6319 [hep-lat]}}.

\bibitem{Andersen:2018mau}
C.~Andersen, J.~Bulava, B.~H\"orz, and C.~Morningstar
  \href{http://dx.doi.org/10.1016/j.nuclphysb.2018.12.018}{{\em Nucl. Phys. B}
  {\bfseries 939} (2019) 145--173},
  \href{http://arxiv.org/abs/1808.05007}{{\ttfamily arXiv:1808.05007
  [hep-lat]}}.

\bibitem{Briceno:2015dca}
R.~A. Briceno, J.~J. Dudek, R.~G. Edwards, C.~J. Shultz, C.~E. Thomas, and
  D.~J. Wilson \href{http://dx.doi.org/10.1103/PhysRevLett.115.242001}{{\em
  Phys. Rev. Lett.} {\bfseries 115} (2015) 242001},
  \href{http://arxiv.org/abs/1507.06622}{{\ttfamily arXiv:1507.06622
  [hep-ph]}}.

\bibitem{Briceno:2016kkp}
R.~A. Brice\~no, J.~J. Dudek, R.~G. Edwards, C.~J. Shultz, C.~E. Thomas, and
  D.~J. Wilson \href{http://dx.doi.org/10.1103/PhysRevD.93.114508}{{\em Phys.
  Rev. D} {\bfseries 93} no.~11, (2016) 114508},
  \href{http://arxiv.org/abs/1604.03530}{{\ttfamily arXiv:1604.03530
  [hep-ph]}}. [Erratum: Phys.Rev.D 105, 079902 (2022)].

\bibitem{Ishizuka:2018qbn}
N.~Ishizuka, K.~I. Ishikawa, A.~Ukawa, and T.~Yoshi\'e
  \href{http://dx.doi.org/10.1103/PhysRevD.98.114512}{{\em Phys. Rev. D}
  {\bfseries 98} no.~11, (2018) 114512},
  \href{http://arxiv.org/abs/1809.03893}{{\ttfamily arXiv:1809.03893
  [hep-lat]}}.

\bibitem{Alexandrou:2018jbt}
C.~Alexandrou, L.~Leskovec, S.~Meinel, J.~Negele, S.~Paul, M.~Petschlies,
  A.~Pochinsky, G.~Rendon, and S.~Syritsyn
  \href{http://dx.doi.org/10.1103/PhysRevD.98.074502}{{\em Phys. Rev. D}
  {\bfseries 98} no.~7, (2018) 074502},
  \href{http://arxiv.org/abs/1807.08357}{{\ttfamily arXiv:1807.08357
  [hep-lat]}}. [Erratum: Phys.Rev.D 105, 019902 (2022)].

\bibitem{RBC:2020kdj}
{RBC, UKQCD} Collaboration, R.~Abbott {\em et~al.}
  \href{http://dx.doi.org/10.1103/PhysRevD.102.054509}{{\em Phys. Rev. D}
  {\bfseries 102} no.~5, (2020) 054509},
  \href{http://arxiv.org/abs/2004.09440}{{\ttfamily arXiv:2004.09440
  [hep-lat]}}.

\bibitem{Radhakrishnan:2022ubg}
A.~Radhakrishnan, J.~J. Dudek, and R.~G. Edwards
  \href{http://arxiv.org/abs/2208.13755}{{\ttfamily arXiv:2208.13755
  [hep-lat]}}.

\bibitem{Baroni:2018iau}
A.~Baroni, R.~A. Brice\~no, M.~T. Hansen, and F.~G. Ortega-Gama
  \href{http://dx.doi.org/10.1103/PhysRevD.100.034511}{{\em Phys. Rev. D}
  {\bfseries 100} no.~3, (2019) 034511},
  \href{http://arxiv.org/abs/1812.10504}{{\ttfamily arXiv:1812.10504
  [hep-lat]}}.

\bibitem{Briceno:2019nns}
R.~A. Brice\~no, M.~T. Hansen, and A.~W. Jackura
  \href{http://dx.doi.org/10.1103/PhysRevD.100.114505}{{\em Phys. Rev. D}
  {\bfseries 100} no.~11, (2019) 114505},
  \href{http://arxiv.org/abs/1909.10357}{{\ttfamily arXiv:1909.10357
  [hep-lat]}}.

\bibitem{Briceno:2020xxs}
R.~A. Brice\~no, M.~T. Hansen, and A.~W. Jackura
  \href{http://dx.doi.org/10.1103/PhysRevD.101.094508}{{\em Phys. Rev. D}
  {\bfseries 101} no.~9, (2020) 094508},
  \href{http://arxiv.org/abs/2002.00023}{{\ttfamily arXiv:2002.00023
  [hep-lat]}}.

\bibitem{Briceno:2020vgp}
R.~A. Brice\~no, A.~W. Jackura, F.~G. Ortega-Gama, and K.~H. Sherman
  \href{http://dx.doi.org/10.1103/PhysRevD.103.114512}{{\em Phys. Rev. D}
  {\bfseries 103} no.~11, (2021) 114512},
  \href{http://arxiv.org/abs/2012.13338}{{\ttfamily arXiv:2012.13338
  [hep-lat]}}.

\bibitem{Lozano:2022kfz}
J.~Lozano, U.-G. Mei\ss{}ner, F.~Romero-L\'opez, A.~Rusetsky, and G.~Schierholz
  \href{http://dx.doi.org/10.1007/JHEP10(2022)106}{{\em JHEP} {\bfseries 10}
  (2022) 106}, \href{http://arxiv.org/abs/2205.11316}{{\ttfamily
  arXiv:2205.11316 [hep-lat]}}.

\bibitem{delaParra:2020rct}
J.~L. de~la Parra, A.~Agadjanov, J.~Gegelia, U.~G. Mei\ss{}ner, and A.~Rusetsky
  \href{http://dx.doi.org/10.22323/1.396.0307}{{\em Phys. Rev. D} {\bfseries
  103} no.~3, (2021) 034507}, \href{http://arxiv.org/abs/2010.10917}{{\ttfamily
  arXiv:2010.10917 [hep-lat]}}.

\bibitem{Briceno:2019opb}
R.~A. Brice\~no, Z.~Davoudi, M.~T. Hansen, M.~R. Schindler, and A.~Baroni
  \href{http://dx.doi.org/10.1103/PhysRevD.101.014509}{{\em Phys. Rev. D}
  {\bfseries 101} no.~1, (2020) 014509},
  \href{http://arxiv.org/abs/1911.04036}{{\ttfamily arXiv:1911.04036
  [hep-lat]}}.

\bibitem{Briceno:2022omu}
R.~A. Brice\~no, A.~W. Jackura, A.~Rodas, and J.~V. Guerrero
  \href{http://arxiv.org/abs/2210.08051}{{\ttfamily arXiv:2210.08051
  [hep-lat]}}.

\bibitem{Sherman:2022tco}
K.~H. Sherman, F.~G. Ortega-Gama, R.~A. Brice\~no, and A.~W. Jackura
  \href{http://dx.doi.org/10.1103/PhysRevD.105.114510}{{\em Phys. Rev. D}
  {\bfseries 105} no.~11, (2022) 114510},
  \href{http://arxiv.org/abs/2202.02284}{{\ttfamily arXiv:2202.02284
  [hep-lat]}}.

\bibitem{Hansen:2019nir}
M.~T. Hansen and S.~R. Sharpe
  \href{http://dx.doi.org/10.1146/annurev-nucl-101918-023723}{{\em Ann. Rev.
  Nucl. Part. Sci.} {\bfseries 69} (2019) 65--107},
\href{http://arxiv.org/abs/1901.00483}{{\ttfamily arXiv:1901.00483 [hep-lat]}}.
%%CITATION = ARXIV:1901.00483;%%.

\bibitem{Rusetsky:2019gyk}
A.~Rusetsky \href{http://dx.doi.org/10.22323/1.363.0281}{{\em PoS} {\bfseries
  LATTICE2019} (2019) 281}, \href{http://arxiv.org/abs/1911.01253}{{\ttfamily
  arXiv:1911.01253 [hep-lat]}}.

\bibitem{Mai:2021lwb}
M.~Mai, M.~D\"oring, and A.~Rusetsky
  \href{http://dx.doi.org/10.1140/epjs/s11734-021-00146-5}{{\em Eur. Phys. J.
  ST} {\bfseries 230} no.~6, (2021) 1623--1643},
  \href{http://arxiv.org/abs/2103.00577}{{\ttfamily arXiv:2103.00577
  [hep-lat]}}.

\bibitem{Romero-Lopez:2021zdo}
F.~Romero-L\'opez
  \href{http://dx.doi.org/10.31349/SuplRevMexFis.3.0308003}{{\em Rev. Mex. Fis.
  Suppl.} {\bfseries 3} no.~3, (2022) 0308003},
  \href{http://arxiv.org/abs/2112.05170}{{\ttfamily arXiv:2112.05170
  [hep-lat]}}.

\bibitem{Polejaeva:2012ut}
K.~Polejaeva and A.~Rusetsky
  \href{http://dx.doi.org/10.1140/epja/i2012-12067-8}{{\em Eur.\ Phys.\ J.\ A}
  {\bfseries 48} (2012) 67}, \href{http://arxiv.org/abs/1203.1241}{{\ttfamily
  arXiv:1203.1241 [hep-lat]}}.

\bibitem{Hansen:2014eka}
M.~T. Hansen and S.~R. Sharpe
  \href{http://dx.doi.org/10.1103/PhysRevD.90.116003}{{\em Phys. Rev.}
  {\bfseries D90} no.~11, (2014) 116003},
\href{http://arxiv.org/abs/1408.5933}{{\ttfamily arXiv:1408.5933 [hep-lat]}}.
%%CITATION = ARXIV:1408.5933;%%.

\bibitem{Hansen:2015zga}
M.~T. Hansen and S.~R. Sharpe
  \href{http://dx.doi.org/10.1103/PhysRevD.92.114509}{{\em Phys. Rev.}
  {\bfseries D92} no.~11, (2015) 114509},
\href{http://arxiv.org/abs/1504.04248}{{\ttfamily arXiv:1504.04248 [hep-lat]}}.
%%CITATION = ARXIV:1504.04248;%%.

\bibitem{Hansen:2015zta}
M.~T. Hansen and S.~R. Sharpe
  \href{http://dx.doi.org/10.1103/PhysRevD.93.014506}{{\em Phys. Rev.}
  {\bfseries D93} (2016) 014506},
\href{http://arxiv.org/abs/1509.07929}{{\ttfamily arXiv:1509.07929 [hep-lat]}}.
%%CITATION = ARXIV:1509.07929;%%.

\bibitem{Hansen:2016ync}
M.~T. Hansen and S.~R. Sharpe
  \href{http://dx.doi.org/10.1103/PhysRevD.95.034501}{{\em Phys. Rev.}
  {\bfseries D95} no.~3, (2017) 034501},
\href{http://arxiv.org/abs/1609.04317}{{\ttfamily arXiv:1609.04317 [hep-lat]}}.
%%CITATION = ARXIV:1609.04317;%%.

\bibitem{Briceno:2017tce}
R.~A. Brice\~no, M.~T. Hansen, and S.~R. Sharpe
  \href{http://dx.doi.org/10.1103/PhysRevD.95.074510}{{\em Phys. Rev.}
  {\bfseries D95} no.~7, (2017) 074510},
\href{http://arxiv.org/abs/1701.07465}{{\ttfamily arXiv:1701.07465 [hep-lat]}}.
%%CITATION = ARXIV:1701.07465;%%.

\bibitem{Briceno:2018mlh}
R.~A. Brice\~no, M.~T. Hansen, and S.~R. Sharpe
  \href{http://dx.doi.org/10.1103/PhysRevD.98.014506}{{\em Phys. Rev.}
  {\bfseries D98} no.~1, (2018) 014506},
\href{http://arxiv.org/abs/1803.04169}{{\ttfamily arXiv:1803.04169 [hep-lat]}}.
%%CITATION = ARXIV:1803.04169;%%.

\bibitem{Briceno:2018aml}
R.~A. Brice\~no, M.~T. Hansen, and S.~R. Sharpe
  \href{http://dx.doi.org/10.1103/PhysRevD.99.014516}{{\em Phys. Rev.}
  {\bfseries D99} no.~1, (2019) 014516},
\href{http://arxiv.org/abs/1810.01429}{{\ttfamily arXiv:1810.01429 [hep-lat]}}.
%%CITATION = ARXIV:1810.01429;%%.

\bibitem{Blanton:2019igq}
T.~D. Blanton, F.~Romero-L\'opez, and S.~R. Sharpe
  \href{http://dx.doi.org/10.1007/JHEP03(2019)106}{{\em JHEP} {\bfseries 03}
  (2019) 106},
\href{http://arxiv.org/abs/1901.07095}{{\ttfamily arXiv:1901.07095 [hep-lat]}}.
%%CITATION = ARXIV:1901.07095;%%.

\bibitem{Briceno:2019muc}
R.~A. Brice\~no, M.~T. Hansen, S.~R. Sharpe, and A.~P. Szczepaniak
  \href{http://dx.doi.org/10.1103/PhysRevD.100.054508}{{\em Phys. Rev.}
  {\bfseries D100} no.~5, (2019) 054508},
\href{http://arxiv.org/abs/1905.11188}{{\ttfamily arXiv:1905.11188 [hep-lat]}}.
%%CITATION = ARXIV:1905.11188;%%.

\bibitem{Jackura:2019bmu}
A.~W. Jackura, S.~M. Dawid, C.~Fern\'andez-Ram\'\i{}rez, V.~Mathieu,
  M.~Mikhasenko, A.~Pilloni, S.~R. Sharpe, and A.~P. Szczepaniak
  \href{http://dx.doi.org/10.1103/PhysRevD.100.034508}{{\em Phys. Rev. D}
  {\bfseries 100} no.~3, (2019) 034508},
  \href{http://arxiv.org/abs/1905.12007}{{\ttfamily arXiv:1905.12007
  [hep-ph]}}.

\bibitem{Romero-Lopez:2019qrt}
F.~Romero-L\'opez, S.~R. Sharpe, T.~D. Blanton, R.~A. Brice\~no, and M.~T.
  Hansen \href{http://dx.doi.org/10.1007/JHEP10(2019)007}{{\em JHEP} {\bfseries
  10} (2019) 007},
\href{http://arxiv.org/abs/1908.02411}{{\ttfamily arXiv:1908.02411 [hep-lat]}}.
%%CITATION = ARXIV:1908.02411;%%.

\bibitem{Hansen:2020zhy}
M.~T. Hansen, F.~Romero-L\'opez, and S.~R. Sharpe
  \href{http://dx.doi.org/10.1007/JHEP07(2020)047}{{\em JHEP} {\bfseries 07}
  (2020) 047}, \href{http://arxiv.org/abs/2003.10974}{{\ttfamily
  arXiv:2003.10974 [hep-lat]}}.

\bibitem{Blanton:2020gmf}
T.~D. Blanton and S.~R. Sharpe
  \href{http://dx.doi.org/10.1103/PhysRevD.103.054503}{{\em Phys. Rev. D}
  {\bfseries 103} no.~5, (2021) 054503},
  \href{http://arxiv.org/abs/2011.05520}{{\ttfamily arXiv:2011.05520
  [hep-lat]}}.

\bibitem{Blanton:2020jnm}
T.~D. Blanton and S.~R. Sharpe
  \href{http://dx.doi.org/10.1103/PhysRevD.102.054515}{{\em Phys. Rev. D}
  {\bfseries 102} no.~5, (2020) 054515},
  \href{http://arxiv.org/abs/2007.16190}{{\ttfamily arXiv:2007.16190
  [hep-lat]}}.

\bibitem{Blanton:2020gha}
T.~D. Blanton and S.~R. Sharpe
  \href{http://dx.doi.org/10.1103/PhysRevD.102.054520}{{\em Phys. Rev. D}
  {\bfseries 102} no.~5, (2020) 054520},
  \href{http://arxiv.org/abs/2007.16188}{{\ttfamily arXiv:2007.16188
  [hep-lat]}}.

\bibitem{Hansen:2021ofl}
M.~T. Hansen, F.~Romero-L\'opez, and S.~R. Sharpe
  \href{http://dx.doi.org/10.1007/JHEP04(2021)113}{{\em JHEP} {\bfseries 04}
  (2021) 113}, \href{http://arxiv.org/abs/2101.10246}{{\ttfamily
  arXiv:2101.10246 [hep-lat]}}.

\bibitem{Blanton:2021mih}
T.~D. Blanton and S.~R. Sharpe
  \href{http://dx.doi.org/10.1103/PhysRevD.104.034509}{{\em Phys. Rev. D}
  {\bfseries 104} no.~3, (2021) 034509},
  \href{http://arxiv.org/abs/2105.12094}{{\ttfamily arXiv:2105.12094
  [hep-lat]}}.

\bibitem{Blanton:2021eyf}
T.~D. Blanton, F.~Romero-L\'opez, and S.~R. Sharpe
  \href{http://dx.doi.org/10.1007/JHEP02(2022)098}{{\em JHEP} {\bfseries 02}
  (2022) 098}, \href{http://arxiv.org/abs/2111.12734}{{\ttfamily
  arXiv:2111.12734 [hep-lat]}}.

\bibitem{Hammer:2017uqm}
H.-W. Hammer, J.-Y. Pang, and A.~Rusetsky
  \href{http://dx.doi.org/10.1007/JHEP09(2017)109}{{\em JHEP} {\bfseries 09}
  (2017) 109},
\href{http://arxiv.org/abs/1706.07700}{{\ttfamily arXiv:1706.07700 [hep-lat]}}.
%%CITATION = ARXIV:1706.07700;%%.

\bibitem{Hammer:2017kms}
H.~W. Hammer, J.~Y. Pang, and A.~Rusetsky
  \href{http://dx.doi.org/10.1007/JHEP10(2017)115}{{\em JHEP} {\bfseries 10}
  (2017) 115},
\href{http://arxiv.org/abs/1707.02176}{{\ttfamily arXiv:1707.02176 [hep-lat]}}.
%%CITATION = ARXIV:1707.02176;%%.

\bibitem{Doring:2018xxx}
M.~{D\"{o}ring}, H.~W. Hammer, M.~Mai, J.~Y. Pang, A.~Rusetsky, and J.~Wu
  \href{http://dx.doi.org/10.1103/PhysRevD.97.114508}{{\em Phys. Rev.}
  {\bfseries D97} no.~11, (2018) 114508},
\href{http://arxiv.org/abs/1802.03362}{{\ttfamily arXiv:1802.03362 [hep-lat]}}.
%%CITATION = ARXIV:1802.03362;%%.

\bibitem{Romero-Lopez:2018rcb}
F.~Romero-L\'opez, A.~Rusetsky, and C.~Urbach
  \href{http://dx.doi.org/10.1140/epjc/s10052-018-6325-8}{{\em Eur. Phys. J.}
  {\bfseries C78} no.~10, (2018) 846},
\href{http://arxiv.org/abs/1806.02367}{{\ttfamily arXiv:1806.02367 [hep-lat]}}.
%%CITATION = ARXIV:1806.02367;%%.

\bibitem{Pang:2019dfe}
J.-Y. Pang, J.-J. Wu, H.~W. Hammer, U.-G. Mei{$\ss$}ner, and A.~Rusetsky
  \href{http://dx.doi.org/10.1103/PhysRevD.99.074513}{{\em Phys. Rev.}
  {\bfseries D99} no.~7, (2019) 074513},
\href{http://arxiv.org/abs/1902.01111}{{\ttfamily arXiv:1902.01111 [hep-lat]}}.
%%CITATION = ARXIV:1902.01111;%%.

\bibitem{Romero-Lopez:2020rdq}
F.~Romero-L\'opez, A.~Rusetsky, N.~Schlage, and C.~Urbach
  \href{http://dx.doi.org/10.1007/JHEP02(2021)060}{{\em JHEP} {\bfseries 02}
  (2021) 060}, \href{http://arxiv.org/abs/2010.11715}{{\ttfamily
  arXiv:2010.11715 [hep-lat]}}.

\bibitem{Muller:2020wjo}
F.~M\"{u}ller and A.~Rusetsky
  \href{http://dx.doi.org/10.1007/JHEP03(2021)152}{{\em JHEP} {\bfseries 21}
  (2020) 152}, \href{http://arxiv.org/abs/2012.13957}{{\ttfamily
  arXiv:2012.13957 [hep-lat]}}.

\bibitem{Muller:2020vtt}
F.~M\"{u}ller, A.~Rusetsky, and T.~Yu
  \href{http://dx.doi.org/10.1103/PhysRevD.103.054506}{{\em Phys. Rev. D}
  {\bfseries 103} no.~5, (2021) 054506},
  \href{http://arxiv.org/abs/2011.14178}{{\ttfamily arXiv:2011.14178
  [hep-lat]}}.

\bibitem{Mai:2017bge}
M.~Mai and M.~{D\"{o}ring}
  \href{http://dx.doi.org/10.1140/epja/i2017-12440-1}{{\em Eur. Phys. J.}
  {\bfseries A53} no.~12, (2017) 240},
\href{http://arxiv.org/abs/1709.08222}{{\ttfamily arXiv:1709.08222 [hep-lat]}}.
%%CITATION = ARXIV:1709.08222;%%.

\bibitem{Guo:2017ism}
P.~Guo and V.~Gasparian
  \href{http://dx.doi.org/10.1016/j.physletb.2017.10.009}{{\em Phys. Lett.}
  {\bfseries B774} (2017) 441--445},
\href{http://arxiv.org/abs/1701.00438}{{\ttfamily arXiv:1701.00438 [hep-lat]}}.
%%CITATION = ARXIV:1701.00438;%%.

\bibitem{Klos:2018sen}
P.~Klos, S.~König, H.~W. Hammer, J.~E. Lynn, and A.~Schwenk
  \href{http://dx.doi.org/10.1103/PhysRevC.98.034004}{{\em Phys. Rev.}
  {\bfseries C98} no.~3, (2018) 034004},
\href{http://arxiv.org/abs/1805.02029}{{\ttfamily arXiv:1805.02029 [nucl-th]}}.
%%CITATION = ARXIV:1805.02029;%%.

\bibitem{Guo:2018ibd}
P.~Guo, M.~D\"{o}ring, and A.~P. Szczepaniak
  \href{http://dx.doi.org/10.1103/PhysRevD.98.094502}{{\em Phys. Rev.}
  {\bfseries D98} no.~9, (2018) 094502},
\href{http://arxiv.org/abs/1810.01261}{{\ttfamily arXiv:1810.01261 [hep-lat]}}.
%%CITATION = ARXIV:1810.01261;%%.

\bibitem{Guo:2019hih}
P.~Guo \href{http://dx.doi.org/10.1016/j.physletb.2020.135370}{{\em Phys. Lett.
  B} {\bfseries 804} (2020) 135370},
  \href{http://arxiv.org/abs/1908.08081}{{\ttfamily arXiv:1908.08081
  [hep-lat]}}.

\bibitem{Pang:2020pkl}
J.-Y. Pang, J.-J. Wu, and L.-S. Geng
  \href{http://dx.doi.org/10.1103/PhysRevD.102.114515}{{\em Phys. Rev. D}
  {\bfseries 102} no.~11, (2020) 114515},
  \href{http://arxiv.org/abs/2008.13014}{{\ttfamily arXiv:2008.13014
  [hep-lat]}}.

\bibitem{Jackura:2020bsk}
A.~W. Jackura, R.~A. Brice\~no, S.~M. Dawid, M.~H.~E. Islam, and C.~McCarty
  \href{http://dx.doi.org/10.1103/PhysRevD.104.014507}{{\em Phys. Rev. D}
  {\bfseries 104} no.~1, (2021) 014507},
  \href{http://arxiv.org/abs/2010.09820}{{\ttfamily arXiv:2010.09820
  [hep-lat]}}.

\bibitem{Muller:2021uur}
F.~M\"uller, J.-Y. Pang, A.~Rusetsky, and J.-J. Wu
  \href{http://arxiv.org/abs/2110.09351}{{\ttfamily arXiv:2110.09351
  [hep-lat]}}.

\bibitem{Jackura:2022gib}
A.~W. Jackura \href{http://arxiv.org/abs/2208.10587}{{\ttfamily
  arXiv:2208.10587 [hep-lat]}}.

\bibitem{Mai:2018djl}
M.~Mai and M.~D{\"{o}}ring
  \href{http://dx.doi.org/10.1103/PhysRevLett.122.062503}{{\em Phys. Rev.
  Lett.} {\bfseries 122} no.~6, (2019) 062503},
\href{http://arxiv.org/abs/1807.04746}{{\ttfamily arXiv:1807.04746 [hep-lat]}}.
%%CITATION = ARXIV:1807.04746;%%.

\bibitem{Mai:2019fba}
M.~Mai, M.~D\"{o}ring, C.~Culver, and A.~Alexandru
  \href{http://dx.doi.org/10.1103/PhysRevD.101.054510}{{\em Phys.\ Rev.\ D}
  {\bfseries 101} (2020) 054510},
  \href{http://arxiv.org/abs/1909.05749}{{\ttfamily arXiv:1909.05749
  [hep-lat]}}.

\bibitem{Culver:2019vvu}
C.~Culver, M.~Mai, R.~Brett, A.~Alexandru, and M.~D\"{o}ring
  \href{http://dx.doi.org/10.1103/PhysRevD.101.114507}{{\em Phys. Rev. D}
  {\bfseries 101} no.~11, (2020) 114507},
  \href{http://arxiv.org/abs/1911.09047}{{\ttfamily arXiv:1911.09047
  [hep-lat]}}.

\bibitem{Guo:2020kph}
P.~Guo and B.~Long \href{http://dx.doi.org/10.1103/PhysRevD.101.094510}{{\em
  Phys. Rev. D} {\bfseries 101} no.~9, (2020) 094510},
  \href{http://arxiv.org/abs/2002.09266}{{\ttfamily arXiv:2002.09266
  [hep-lat]}}.

\bibitem{Alexandru:2020xqf}
A.~Alexandru, R.~Brett, C.~Culver, M.~D\"{o}ring, D.~Guo, F.~X. Lee, and M.~Mai
  \href{http://dx.doi.org/10.1103/PhysRevD.102.114523}{{\em Phys. Rev. D}
  {\bfseries 102} no.~11, (2020) 114523},
  \href{http://arxiv.org/abs/2009.12358}{{\ttfamily arXiv:2009.12358
  [hep-lat]}}.

\bibitem{Hansen:2020otl}
{Hadron Spectrum} Collaboration, M.~T. Hansen, R.~A. Brice\~no, R.~G. Edwards,
  C.~E. Thomas, and D.~J. Wilson
  \href{http://dx.doi.org/10.1103/PhysRevLett.126.012001}{{\em Phys. Rev.
  Lett.} {\bfseries 126} (2021) 012001},
  \href{http://arxiv.org/abs/2009.04931}{{\ttfamily arXiv:2009.04931
  [hep-lat]}}.

\bibitem{Brett:2021wyd}
R.~Brett, C.~Culver, M.~Mai, A.~Alexandru, M.~D\"oring, and F.~X. Lee
  \href{http://dx.doi.org/10.1103/PhysRevD.104.014501}{{\em Phys. Rev. D}
  {\bfseries 104} no.~1, (2021) 014501},
  \href{http://arxiv.org/abs/2101.06144}{{\ttfamily arXiv:2101.06144
  [hep-lat]}}.

\bibitem{Mai:2021nul}
{GWQCD} Collaboration, M.~Mai, A.~Alexandru, R.~Brett, C.~Culver, M.~D\"oring,
  F.~X. Lee, and D.~Sadasivan
  \href{http://dx.doi.org/10.1103/PhysRevLett.127.222001}{{\em Phys. Rev.
  Lett.} {\bfseries 127} no.~22, (2021) 222001},
  \href{http://arxiv.org/abs/2107.03973}{{\ttfamily arXiv:2107.03973
  [hep-lat]}}.

\bibitem{Blanton:2021llb}
T.~D. Blanton, A.~D. Hanlon, B.~H\"orz, C.~Morningstar, F.~Romero-L\'opez, and
  S.~R. Sharpe \href{http://dx.doi.org/10.1007/JHEP10(2021)023}{{\em JHEP}
  {\bfseries 10} (2021) 023}, \href{http://arxiv.org/abs/2106.05590}{{\ttfamily
  arXiv:2106.05590 [hep-lat]}}.

\bibitem{HadronSpectrum:2009krc}
{Hadron Spectrum} Collaboration, M.~Peardon, J.~Bulava, J.~Foley,
  C.~Morningstar, J.~Dudek, R.~G. Edwards, B.~Joo, H.-W. Lin, D.~G. Richards,
  and K.~J. Juge \href{http://dx.doi.org/10.1103/PhysRevD.80.054506}{{\em Phys.
  Rev. D} {\bfseries 80} (2009) 054506},
  \href{http://arxiv.org/abs/0905.2160}{{\ttfamily arXiv:0905.2160 [hep-lat]}}.

\bibitem{Horz:2019rrn}
B.~H\"{o}rz and A.~Hanlon
  \href{http://dx.doi.org/10.1103/PhysRevLett.123.142002}{{\em Phys. Rev.
  Lett.} {\bfseries 123} no.~14, (2019) 142002},
\href{http://arxiv.org/abs/1905.04277}{{\ttfamily arXiv:1905.04277 [hep-lat]}}.
%%CITATION = ARXIV:1905.04277;%%.

\bibitem{3neutrons}
Z.~T. Draper, M.~T. Hansen, F.~Romero-L\'opez, and S.~R. Sharpe
\newblock 2022.
\newblock \href{http://arxiv.org/abs/in progress}{{\ttfamily in progress}}.

\bibitem{Garofalo:2022pux}
M.~Garofalo, M.~Mai, F.~Romero-L\'opez, A.~Rusetsky, and C.~Urbach
  \href{http://arxiv.org/abs/2211.05605}{{\ttfamily arXiv:2211.05605
  [hep-lat]}}.

\bibitem{Muller:2022oyw}
F.~M\"uller, J.-Y. Pang, A.~Rusetsky, and J.-J. Wu
  \href{http://arxiv.org/abs/2211.10126}{{\ttfamily arXiv:2211.10126
  [hep-lat]}}.

\bibitem{Bijnens:2021hpq}
J.~Bijnens and T.~Husek
  \href{http://dx.doi.org/10.1103/PhysRevD.104.054046}{{\em Phys. Rev. D}
  {\bfseries 104} no.~5, (2021) 054046},
  \href{http://arxiv.org/abs/2107.06291}{{\ttfamily arXiv:2107.06291
  [hep-ph]}}.

\bibitem{Bijnens:2022zsq}
J.~Bijnens, T.~Husek, and M.~Sj\"o
  \href{http://dx.doi.org/10.1103/PhysRevD.106.054021}{{\em Phys. Rev. D}
  {\bfseries 106} no.~5, (2022) 054021},
  \href{http://arxiv.org/abs/2206.14212}{{\ttfamily arXiv:2206.14212
  [hep-ph]}}.

\bibitem{QC3git}
T.~Blanton, F.~Romero-L\'opez, and S.~Sharpe.
  \url{https://github.com/ferolo2/QC3_release}.
  https://github.com/ferolo2/QC3\_release.

\bibitem{Beane:2007qr}
S.~R. Beane, W.~Detmold, and M.~J. Savage
  \href{http://dx.doi.org/10.1103/PhysRevD.76.074507}{{\em Phys. Rev. D}
  {\bfseries 76} (2007) 074507},
  \href{http://arxiv.org/abs/0707.1670}{{\ttfamily arXiv:0707.1670 [hep-lat]}}.

\bibitem{Detmold:2008fn}
W.~Detmold, M.~J. Savage, A.~Torok, S.~R. Beane, T.~C. Luu, K.~Orginos, and
  A.~Parreno \href{http://dx.doi.org/10.1103/PhysRevD.78.014507}{{\em Phys.
  Rev.} {\bfseries D78} (2008) 014507},
\href{http://arxiv.org/abs/0803.2728}{{\ttfamily arXiv:0803.2728 [hep-lat]}}.
%%CITATION = ARXIV:0803.2728;%%.

\bibitem{Detmold:2010au}
W.~Detmold and M.~J. Savage
  \href{http://dx.doi.org/10.1103/PhysRevD.82.014511}{{\em Phys. Rev. D}
  {\bfseries 82} (2010) 014511},
  \href{http://arxiv.org/abs/1001.2768}{{\ttfamily arXiv:1001.2768 [hep-lat]}}.

\bibitem{Detmold:2011kw}
W.~Detmold and B.~Smigielski
  \href{http://dx.doi.org/10.1103/PhysRevD.84.014508}{{\em Phys. Rev. D}
  {\bfseries 84} (2011) 014508},
  \href{http://arxiv.org/abs/1103.4362}{{\ttfamily arXiv:1103.4362 [hep-lat]}}.

\bibitem{Detmold:2012pi}
W.~Detmold, S.~Meinel, and Z.~Shi
  \href{http://dx.doi.org/10.1103/PhysRevD.87.094504}{{\em Phys. Rev. D}
  {\bfseries 87} no.~9, (2013) 094504},
  \href{http://arxiv.org/abs/1211.3156}{{\ttfamily arXiv:1211.3156 [hep-lat]}}.

\bibitem{Detmold:2014iha}
W.~Detmold \href{http://dx.doi.org/10.1103/PhysRevLett.114.222001}{{\em Phys.
  Rev. Lett.} {\bfseries 114} no.~22, (2015) 222001},
  \href{http://arxiv.org/abs/1408.6919}{{\ttfamily arXiv:1408.6919 [hep-ph]}}.

\bibitem{manypi}
R.~Abbott, W.~Detmold, and F.~Romero-L\'opez
\newblock 2022.
\newblock \href{http://arxiv.org/abs/in progress}{{\ttfamily in progress}}.

\bibitem{Endres:2011jm}
M.~G. Endres, D.~B. Kaplan, J.-W. Lee, and A.~N. Nicholson
  \href{http://dx.doi.org/10.1103/PhysRevLett.107.201601}{{\em Phys. Rev.
  Lett.} {\bfseries 107} (2011) 201601},
  \href{http://arxiv.org/abs/1106.0073}{{\ttfamily arXiv:1106.0073 [hep-lat]}}.

\bibitem{Yunus:2022wuc}
C.~Yunus and W.~Detmold \href{http://arxiv.org/abs/2210.15789}{{\ttfamily
  arXiv:2210.15789 [hep-lat]}}.

\bibitem{Hansen:2019idp}
M.~Hansen, A.~Lupo, and N.~Tantalo
  \href{http://dx.doi.org/10.1103/PhysRevD.99.094508}{{\em Phys. Rev. D}
  {\bfseries 99} no.~9, (2019) 094508},
  \href{http://arxiv.org/abs/1903.06476}{{\ttfamily arXiv:1903.06476
  [hep-lat]}}.

\bibitem{Bulava:2019kbi}
J.~Bulava and M.~T. Hansen
  \href{http://dx.doi.org/10.1103/PhysRevD.100.034521}{{\em Phys. Rev. D}
  {\bfseries 100} no.~3, (2019) 034521},
  \href{http://arxiv.org/abs/1903.11735}{{\ttfamily arXiv:1903.11735
  [hep-lat]}}.

\bibitem{Bulava:2021fre}
J.~Bulava, M.~T. Hansen, M.~W. Hansen, A.~Patella, and N.~Tantalo
  \href{http://dx.doi.org/10.1007/JHEP07(2022)034}{{\em JHEP} {\bfseries 07}
  (2022) 034}, \href{http://arxiv.org/abs/2111.12774}{{\ttfamily
  arXiv:2111.12774 [hep-lat]}}.

\bibitem{Bruno:2020kyl}
M.~Bruno and M.~T. Hansen \href{http://dx.doi.org/10.1007/JHEP06(2021)043}{{\em
  JHEP} {\bfseries 06} (2021) 043},
  \href{http://arxiv.org/abs/2012.11488}{{\ttfamily arXiv:2012.11488
  [hep-lat]}}.

\bibitem{Garofalo:2021bzl}
M.~Garofalo, F.~Romero-L\'opez, A.~Rusetsky, and C.~Urbach
  \href{http://dx.doi.org/10.1140/epjc/s10052-021-09830-1}{{\em Eur. Phys. J.
  C} {\bfseries 81} no.~11, (2021) 1034},
  \href{http://arxiv.org/abs/2107.04853}{{\ttfamily arXiv:2107.04853
  [hep-lat]}}.

\bibitem{Detmold:2004qn}
W.~Detmold and M.~J. Savage
  \href{http://dx.doi.org/10.1016/j.nuclphysa.2004.07.007}{{\em Nucl. Phys. A}
  {\bfseries 743} (2004) 170--193},
  \href{http://arxiv.org/abs/hep-lat/0403005}{{\ttfamily
  arXiv:hep-lat/0403005}}.

\bibitem{Beane:2012ey}
S.~R. Beane, E.~Chang, S.~D. Cohen, W.~Detmold, H.~W. Lin, T.~C. Luu,
  K.~Orginos, A.~Parreno, M.~J. Savage, and A.~Walker-Loud
  \href{http://dx.doi.org/10.1103/PhysRevLett.109.172001}{{\em Phys. Rev.
  Lett.} {\bfseries 109} (2012) 172001},
  \href{http://arxiv.org/abs/1204.3606}{{\ttfamily arXiv:1204.3606 [hep-lat]}}.

\bibitem{Li:2021mob}
Y.~Li, J.-j. Wu, D.~B. Leinweber, and A.~W. Thomas
  \href{http://dx.doi.org/10.1103/PhysRevD.103.094518}{{\em Phys. Rev. D}
  {\bfseries 103} no.~9, (2021) 094518},
  \href{http://arxiv.org/abs/2103.12260}{{\ttfamily arXiv:2103.12260
  [hep-lat]}}.

\bibitem{Li:2019qvh}
Y.~Li, J.-J. Wu, C.~D. Abell, D.~B. Leinweber, and A.~W. Thomas
  \href{http://dx.doi.org/10.1103/PhysRevD.101.114501}{{\em Phys. Rev. D}
  {\bfseries 101} no.~11, (2020) 114501},
  \href{http://arxiv.org/abs/1910.04973}{{\ttfamily arXiv:1910.04973
  [hep-lat]}}.

\bibitem{Severt:2022jtg}
D.~Severt, M.~Mai, and U.-G. Mei\ss{}ner
  \href{http://arxiv.org/abs/2212.02171}{{\ttfamily arXiv:2212.02171
  [hep-lat]}}.

\bibitem{Eliyahu:2019nkz}
M.~Eliyahu, B.~Bazak, and N.~Barnea
  \href{http://dx.doi.org/10.1103/PhysRevC.102.044003}{{\em Phys. Rev. C}
  {\bfseries 102} no.~4, (2020) 044003},
  \href{http://arxiv.org/abs/1912.07017}{{\ttfamily arXiv:1912.07017
  [nucl-th]}}.

\bibitem{Detmold:2021oro}
W.~Detmold and P.~E. Shanahan
  \href{http://dx.doi.org/10.1103/PhysRevD.103.074503}{{\em Phys. Rev. D}
  {\bfseries 103} no.~7, (2021) 074503},
  \href{http://arxiv.org/abs/2102.04329}{{\ttfamily arXiv:2102.04329
  [nucl-th]}}.

\bibitem{Sun:2022frr}
X.~Sun, W.~Detmold, D.~Luo, and P.~E. Shanahan
  \href{http://dx.doi.org/10.1103/PhysRevD.105.074508}{{\em Phys. Rev. D}
  {\bfseries 105} no.~7, (2022) 074508},
  \href{http://arxiv.org/abs/2202.03530}{{\ttfamily arXiv:2202.03530
  [nucl-th]}}.

\bibitem{NPLQCD:2012mex}
{NPLQCD} Collaboration, S.~R. Beane, E.~Chang, S.~D. Cohen, W.~Detmold, H.~W.
  Lin, T.~C. Luu, K.~Orginos, A.~Parreno, M.~J. Savage, and A.~Walker-Loud
  \href{http://dx.doi.org/10.1103/PhysRevD.87.034506}{{\em Phys. Rev. D}
  {\bfseries 87} no.~3, (2013) 034506},
  \href{http://arxiv.org/abs/1206.5219}{{\ttfamily arXiv:1206.5219 [hep-lat]}}.

\end{thebibliography}\endgroup

\end{document}